\documentclass{aastex631}

\definecolor{brightviolet}{rgb}{0.8, 0.4, 1}
\newcommand{\q}[1]{\textcolor{brightviolet}{#1}}

\usepackage{soul,xcolor}
\usepackage{graphicx}
\usepackage{amsmath}
\setstcolor{red}
\usepackage{multirow}

\begin{document}

\title{Pulsar scattering as a probe for structures in the interstellar medium}

\author[0009-0009-1205-7294]{Qiuyi He}
\affiliation{South-Western Institute for Astronomy Research (SWIFAR) \\ Yunnan University, 650500 Kunming, P. R. China}

\author[0000-0003-2076-4510]{Xun Shi}
\affiliation{South-Western Institute for Astronomy Research (SWIFAR) \\ Yunnan University, 650500 Kunming, P. R. China}

\author[0000-0003-3144-1952]{Guangxing Li}
\affiliation{South-Western Institute for Astronomy Research (SWIFAR) \\ Yunnan University, 650500 Kunming, P. R. China}



\begin{abstract}

Due to the inhomogeneity of electron number density, radio waves emitted by pulsars undergo scattering as they pass through the interstellar medium (ISM). However, a connection between large-scale pulsar scattering data and the structure of the Galactic ISM has yet to be established. In this paper, we explore the capability of pulsar scattering time data in discovering structures in the ISM.  Using a large dataset of scattering time measurements for 473 pulsars, we fit the pulsar reduced scattering intensity as a function of Galactic latitude and distance, constructing a smooth model of the Galactic pulsar scattering distribution. By comparing this smooth distribution with observational data, 
we identify two ISM structures responsible for pulsar scattering, one is associated with the Vela supernova remnant region within the Gum Nebula, while the other is a newly discovered structure -- a distant superbubble, G38, located at a distance of 2.3\,kpc with a size of $\sim$50\,pc. Analysis of the correlation coefficient of the pulsar scattering distribution shows that the correlation is dominated by structures smaller than 0.15\,kpc -- the closest separation approachable by the current dataset.
As measurements of the pulsar scattering time continue to increase in the future, they can potentially become an independent tool for exploring structures in the ISM.

\end{abstract}



\section{Introduction} 
\label{Intro}

Pulsars, renowned for their precise periodic radio emissions, are often referred to as ``cosmic lighthouses.'' As pulsar radio signals propagate through the interstellar medium (ISM), they are affected by several propagation effects. Dispersion, caused by the different propagation speeds of electromagnetic waves at different frequencies through the ISM, leads to delays in the arrival times of signals, which is quantified by the dispersion measure \citep[DM;][]{dispersion1, dispersion2}. Due to the small-scale structures and turbulence in the ionized ISM, signals undergo deflections during their propagation, leading to a broadening of both the observed image and the pulse profile \citep{Scheuer68, Rickett70, Sutton71}. The same multi-path propagation also leads to scintillation i.e. rapid fluctuations in signal intensity \citep{scint1, scint2}.

Numerous research findings indicate that the propagation effects of pulsar signals provide a means to study structures within the Galaxy. \citet{HII69} discovered that the ${\rm H}\,_{\rm II}$ region around the star will have some effect on the dispersion of Galactic pulsars. \citet{HI} investigated the distribution of ${\rm H}\,_{\rm I}$ regions along the line of sight by comparing pulsar ${\rm H}\,_{\rm I}$ absorption measurements at different times. \citet{Davidson69} through the observation of pulsars at different frequencies, calculated the electron number density along the path between the pulsar and the observer using the relationship between arrival time and frequency, and they also considered the impact of ionization from early-type stars on the surrounding ISM in the context of pulsar scattering. Subsequently, the study of pulsar scintillation was significantly advanced by observations of parabolic arc and inverted arclet structures in pulsar secondary spectra, as reported by \citet{Stinebring01}, \citet{Hill03, Hill05}, \citet{Brisken10}. These observational findings were later explained and further developed through theoretical works \citep[e.g.][]{Walker04, Cordes06, Pen14}. These studies not only enhance our understanding of the structures within the Galaxy but also reveal the potential value of pulsar observations in deciphering the complex structures of the ISM \citep{SX, Baker23, Ocker2, Ocker}. 

Particularly, pulsar observations have enabled the construction of Galactic electron density models. The initial exploration was initiated by \citet{Manchester81} and \citet{Lyne85}. They proposed the LMT85 Galaxy model based on the motion distances obtained from ${\rm H}\,_{\rm I}$ absorption of 36 pulsars. Subsequently, \citet{Taylor93} described the quantitative model TC93 for the distribution of free electrons in the Galaxy in 1993. This was the first consideration of observational results related to interstellar scattering, emphasizing the practicality of estimating pulsar distances through dispersion measurements. Since 2002, the NE2001 model proposed by \citet{Cordes02, Cordes03} has become the standard for estimating pulsar distances. It takes into account large-scale variations in neon intensity caused by interstellar scattering, local hot bubbles around the Sun, superbubbles, as well as structures like nebulae and supernova remnants. Based on the NE2001 model, \citet{YMW16} developed a new Galactic electron density model YMW16, by incorporating more structures and utilizing a larger dataset.

Multiwavelength observations toward the interstellar gas of the Galaxy have accumulated a wealth of data, providing insights into the structure of dust and gas within the Galaxy. Recent highlights include the Planck map \citep{Plank19, Plank192} tracing the spatial distribution of dust and gas in the sky, 3D dust maps \citep[e.g.][]{dustmap} which include valuable distance information, full sky ${\rm H}\,_{\rm I}$ data \citep{HI4PI} tracing the distribution of warm ionized gas in the position-position-velocity space, the Wide-field Infrared Survey Explorer (WISE) catalog of ${\rm H}\,_{\rm II}$ regions \citep{WISE}, and high-quality radio recombination lines (RRLs) data tracing ionized gas at different scales \citep{RRL1, RRL2}.

Pulsar scattering is caused by fluctuations in electron density within the ISM, and it can be quantified by measuring the temporal broadening of pulsar signals. %Large-scale 
Macroscopic astrophysical structures can significantly influence the scattering of pulsar signals.
Supernova remnants \citep{cxx1, cxx2}, local interstellar clouds \citep{B1133}, ionized skins of molecular clumps \citep{walker17}, ${\rm H}\,_{\rm I}$ filaments \citep{Stock24}, and ${\rm H}\,_{\rm II}$ regions %, which 
have long been considered feasible explanations for enhanced DM and scattering in the inner Galaxy \citep{HII01, HII02, Ocker24}.
These structures can potentially leave coherent patterns on pulsar scattering.
We look for such coherent patterns in the 3D distribution of pulsar reduced scattering intensity in the Galaxy, and cross-correlate this distribution with other ISM data for confirmation and identification of the foreground structure. This way, we explore the ability of pulsar scattering data in exploring ISM structures, and how it can be integrated into the multiwavelength view of the ISM. 

%We discover two coherent patterns in the distribution of pulsar scattering. One is related to the nearby Gum nebula which has been known to have a significant effect on pulsar scattering. The other we identify as a distant superbubble G38. With the pulsar scattering data, we fit the pulsar scattering strength as a smooth function of Galactic latitude and distance. Strong scattering caused by large astrophysical structures should stand out after subtracting this smooth function, as G38 does. With more high-quality pulsar scattering data coming up, we expect more ISM structures to be detected this way.  

%This study explores the ability of pulsar scattering data in exploring ISM structures, and how it can be integrated into the multiwavelength view of the ISM. 
%\hqy{In this paper, we utilize a pulsar dataset with scattering time measurements to explore pulsars' potential for studying ISM structures in the Galaxy.} 
We shall introduce the data in Section~\ref{Data}, fit the smooth distribution of pulsar scattering, and analyze the correlation coefficient of pulsar scattering distribution in Section~\ref{Scattering_fits}, present the coherent scattering patterns caused by the Vela supernova remnant and G38 regions in Section~\ref{Results}.

\section{Data collection and processing}

\subsection{pulsar scattering time data}
\label{Data}

We use a dataset with 473 pulsar scattering time measurements that we previously compiled based on the ATNF catalog \footnote{\url{https://www.atnf.csiro.au/people/pulsar/psrcat/}} 
\citep{ATNF}, which is the largest multifrequency pulsar scattering time dataset to date (see Appendix B in \citet{He24} for details). 
%In this paper, we utilize this dataset to explore its potential for studying ISM structures in the Galaxy. 
For pulsars with multifrequency pulsar scattering measurements, we scale the measurements to 1\,GHz using the $\tau \propto \nu^4$ scaling and take the average of all measurements. Here $\nu$ is the observing frequency.
%Given that the collected data includes multifrequency observations for the same pulsars, we take the average of the reduced scattering intensity for pulsars with multifrequency results. 
We shall study the distribution of the reduced scattering intensity $\tilde{\tau}=\frac{\tau \cdot d}{{\rm DM}^{2}}\sim \frac{\left \langle {\mathrm{\Delta}} {n_{\rm e}}^{2}\right \rangle}{\left \langle n_{\rm e}\right \rangle^{2}}$, where $d$ represents the pulsar's distance along the line of sight, dispersion measure $\mathrm{DM}\propto \int_{0}^{d} n_{\rm e}\rm{d} \ell = \mathit{\left \langle n_{\rm e}\right \rangle d}$, and scattering time $\tau \propto \int_{0}^{d} \left (n_{\rm e}^2 -\bar{n}_{e} ^{2} \right ) \rm{d} \ell = \mathit{ \left\langle {\mathrm{\Delta}} n_{e}^2 \right\rangle d}$. The reduced scattering intensity was introduced in \citet{He24} to have no apparent dependence on the pulsar distance or the average electron number density. It directly reflects the reduced fluctuations in the electron density and can be related to the fluctuation parameters in the Galactic scattering model. 

\begin{figure}
    \centering
	\includegraphics[width=\linewidth]{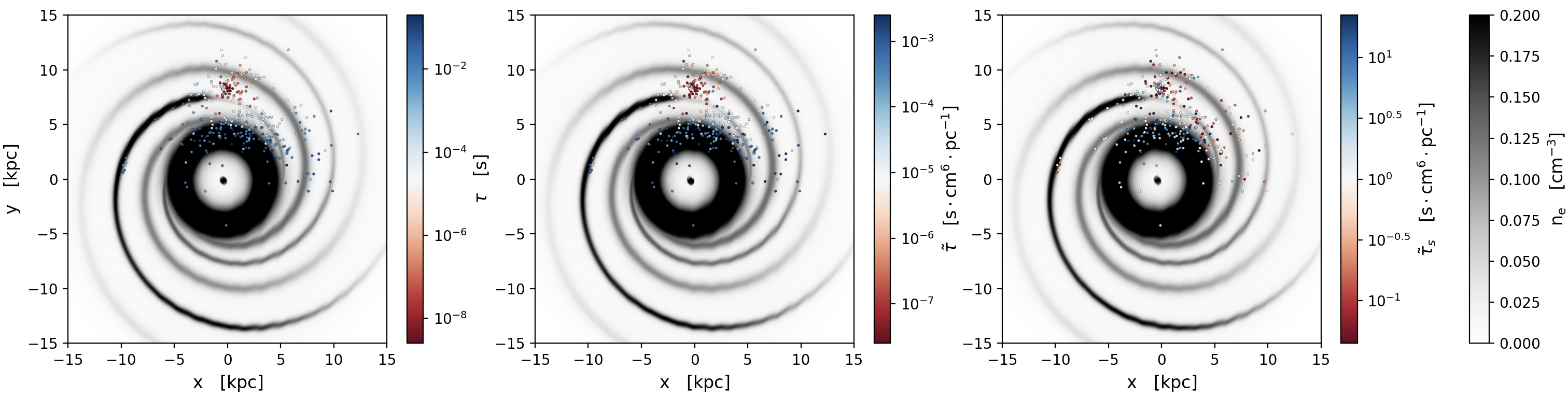}
    \caption{\textbf{Distribution of pulsar scattering intensity (left), reduced scattering intensity (middle), and fluctuation scattering intensity (right) for pulsars with distance from the Galactic plane less than 0.4\,kpc.} 
    The background is the electron density model YMW16 provided by \citet{YMW16}. The range of the reduced scattering intensity introduced by \citet{He24} is smaller than that of the simple scattering intensity, and the fluctuation scattering intensity is distributed around zero.}
    \label{fig:tau_tautl_tautls}
\end{figure}

Fig.~\ref{fig:tau_tautl_tautls} shows the Galactic distribution of pulsars in our dataset with $|z| < 0.4$\,kpc, with $z$ being the distance from the Galactic plane. In the left and the middle panel, the color of the data points indicates the values of the pulsar scattering time ${\tau}$ and the reduced pulsar scattering strength $\tilde{\tau}$, respectively. Although both ${\tau}$ and $\tilde{\tau}$ tend to have lower values in the local region and higher values in the inner Galaxy, their implications are different. With the apparent dependencies on distance and mean electron density subtracted, the values of $\tilde{\tau}$ span a much smaller range compared to those of ${\tau}$. That the $\tilde{\tau}$ values are higher in the inner Galaxy directly reflects higher electron density fluctuations there.

%where $D$ represents the pulsar's distance along the line of sight, dispersion measure $\mathrm{DM}\propto \int_{0}^{D} n_{\rm e}\rm{d} \ell = \mathit{\left \langle n_{\rm e}\right \rangle D}$, and scattering time $\tau \propto \int_{0}^{D} \left (n_{\rm e}^2 -\bar{ n_{e}} ^{2} \right ) \rm{d} \ell = \mathit{ \left\langle {\mathrm{\Delta}} n_{e}^2 \right\rangle D}$. 

%The left panel shows the scattering intensity $\tau$, ranging from $10^{-9.3}$ to $10^{0}$ s. The middle panel displays the reduced scattering intensity $\tilde{\tau}$, with a smaller range of $10^{-7.6}$ to $10^{-2.6}$\,s\,cm$^{6}$\,pc$^{-1}$. The right panel displays the fluctuation scattering intensity $\tilde{\tau}_s$, which is evenly distributed around zero. The definition of the fluctuation scattering intensity is discussed in detail in Section~\ref{Scattering_fits}.}

%Fig.~\ref{fig:tau_tautl_tautls} presents the Galactic distribution of the reduced scattering intensity for all pulsars in our dataset. This visualization reveals potential relationships between pulsar scattering and the structural features of the Galaxy\hqy{, which will be analyzed in detail in Section~\ref{se:correlation}.}
%For the convenience of subsequent data processing and simulation analysis, and to position the center of the Galaxy at the middle, we subtract $360^{\circ}$ from the longitude of pulsars with $l>180^{\circ}$, thus adjusting their range to $-180^{\circ}<l<0^{\circ}$.

\subsection{Statistical Pattern of Pulsar Scattering in the Galaxy}
\label{Scattering_fits}

\subsubsection{Fitting formula for the smooth distribution}
\label{fitting}

\begin{figure}
    \centering
	\includegraphics[width=10cm]{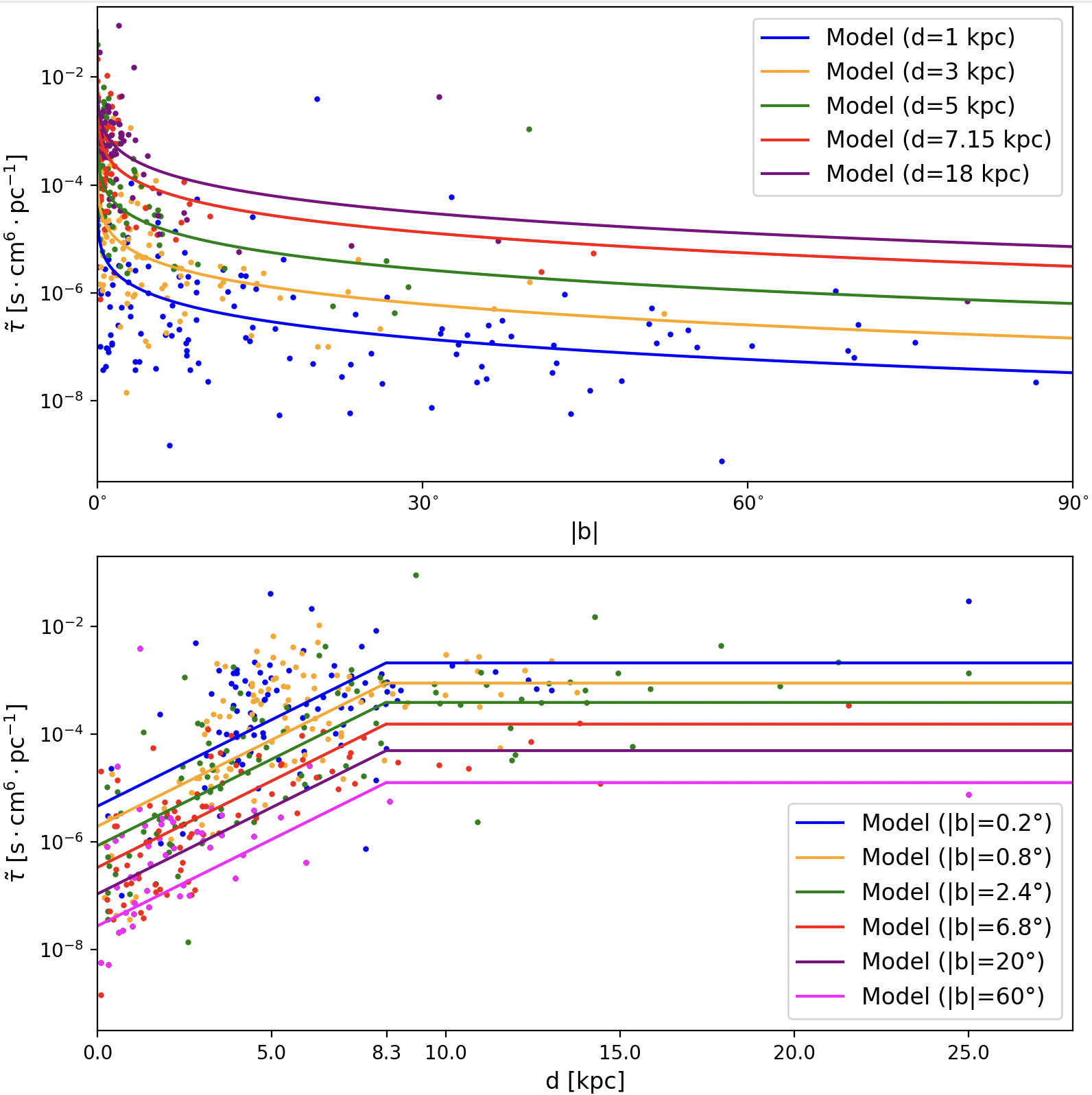}
    \caption{\textbf{The reduced scattering intensity vs. the latitude (top) and distance (bottom) of the pulsar from the Galaxy.} 
    (Top) Pulsars are grouped by distance into five bins: $d < 2$\,kpc, 2\,kpc $\leq d < 4$\,kpc, 4\,kpc $\leq d < 6$\,kpc, 6\,kpc $\leq d < 8.3$\,kpc, and $d \geq 8.3$\,kpc. The groups are represented by the colors `blue,' `orange,' `green,' `red,' and `purple,' respectively. Model lines are plotted at $d = 1$\,kpc, $3$\,kpc, $5$\,kpc, $7.15$\,kpc, and $18$\,kpc, illustrating the $\tilde{\tau}$ - $|b|$ relationship for each group.
    (Bottom) Pulsars are grouped by Galactic latitude $|b|$ into six bins: $|b| < 0.4^\circ$, $0.4^\circ \leq |b| < 1.2^\circ$, $1.2^\circ \leq |b| < 3.6^\circ$, $3.6^\circ \leq |b| < 10^\circ$, $10^\circ \leq |b| < 30^\circ$, and $|b| \geq 30^\circ$. The groups are represented by the colors `blue', `orange,' `green,' `red,' `purple,' and `magenta,' respectively. Model lines are plotted at $|b| = 0.2^\circ$, $0.8^\circ$, $2.4^\circ$, $6.8^\circ$, $20^\circ$, and $60^\circ$, showing the $\tilde{\tau}$ - $d$ relationship for each group.}
    \label{fig:tau_b_d}
\end{figure}

%\hqy{Density fluctuations %in large-scale astrophysical structures within the interstellar medium can significantly impact the scattering intensity of pulsar signals passing through them \citep{HII01, HII02, walker17, Ocker24}.} 
In our Galaxy, the distribution of pulsar scattering intensity exhibits some large-scale trend: pulsar scattering, even in terms of reduced scattering intensity, tends to increase toward the inner Galaxy and decrease with Galactic latitude. Fitting and removing these smooth dependencies would allow one to better discover coherent patterns in pulsar scattering on smaller scales and to better evaluate their significance.
%To further investigate whether this local scattering enhancement is real and its significance, and to discover less-pronounced coherent patterns in pulsar scattering, we must first remove the smooth dependence of pulsar reduced scattering strength on its location. %position. 
%Through interpolating the data (see the bottom of Fig.~\ref{fig:RRL}), we observe that scattering strength shows no obvious correlation with Galactic longitude ($l$). 
Through the analysis of the 3D distribution of reduced scattering intensity for all pulsars in the dataset, we find no significant correlation with Galactic longitude ($l$). In contrast, prominent correlations are observed with Galactic latitude ($b$) and distance ($d$) (see Fig.~\ref{fig:tau_b_d}). Since scattering strength for different pulsars often differs by orders of magnitude, we use the logarithm of the reduced scattering strength, $\log {\tilde{\tau}}$, as our variable here. %The reduced scattering strength 
$\log {\tilde{\tau}}$ follows a simple power-law relationship with latitude, which we describe using the formula $\log {\tilde{\tau}(b)} = {\rm A} \times \left | b \right | ^{\rm a} + {\rm C_{1}}$, where $b$ has a unit of a degree. % where $\rm A$ and $\rm a$ are the coefficient and exponent of the power-law relationship, and $\rm C_{1}$ is an offset term that adjusts the vertical position of the curve to reflect the overall trend of the reduced scattering intensity distribution. 
On the other hand, the relationship between $\log {\tilde{\tau}}$ and distance shows a piecewise characteristic: it exhibits linear growth when the distance is less than 8.3\,kpc, i.e. for a pulsar closer to the Galactic center, but flattens beyond that.
Therefore, we use the Heaviside step function $H(x)$ and write the dependence on $d$ as $\log {\tilde{\tau}(d)} = {\rm B} \times d \times H(8.3-d) + {\rm B} \times 8.3 \times H(d -8.3) +  {\rm C_{2}}$, where %$\rm B$ is the coefficient representing the linear growth of reduced scattering intensity with distance, and $\rm C_{2}$ is a baseline term that characterizes the overall distribution of reduced scattering intensity as a function of distance.
$d$ is in units of kiloparsec.

%\hqy{The relationship between Galactic latitude ($b$) and distance ($d$) with the reduced scattering intensity ($\tilde{\tau}$) acts simultaneously, and we assume their effects are multiplicative: $\tilde{\tau} = \tilde{\tau}(b) \cdot \tilde{\tau}(d)$. Based on this assumption, 
Combining these two dependencies, we adopt the following parameterized form for fitting:
\begin{equation}
\log {\tilde{\tau}} = {\rm A} \times \left | b \right | ^{\rm a} +  {\rm B} \times d \times H(8.3-d) + {\rm B} \times 8.3 \times H(d -8.3) +  {\rm C}
\label{eq:fitting}
\end{equation}
where ${\rm C} = {\rm C_{1}} + {\rm C_{2}}$. 
We employ the Markov Chain Monte Carlo (MCMC) method to fit the parameters \({\rm A}, {\rm a}, {\rm B}, {\rm C}\) of the reduced scattering intensity model, simultaneously capturing the combined effects of Galactic latitude and distance on the reduced scattering intensity. The posterior distributions of the parameters are presented in Appendix~\ref{ap:mcmc}.
The final best-fit values for the model parameters are as follows:
\begin{equation}
\begin{matrix}
{\rm A} &= -1.99^{+0.086}_{-0.093}, \quad {\rm a} = 0.156^{+0.006}_{-0.006}, \\
{\rm B} &= 0.321^{+0.002}_{-0.002}, \quad {\rm C} = -3.79^{+0.092}_{-0.085}.\nonumber
\end{matrix}
\end{equation}

The contribution to the reduced scattering strength $\tilde{\tau}$ of pulsars can be divided into two parts: one is the contribution from the smooth, large-scale scattering of the Galaxy $\tilde{\tau}_{\rm g}$, the other is the fluctuation scattering intensity $\tilde{\tau}_{\rm s}$.
%and the other is the contribution from local interstellar structures affecting pulsar scattering $\tilde{\tau}_{\rm s}$.
The above fitting gives the smooth part:
%We obtained a smooth function of pulsar scattering intensity as a function of Galactic latitude and distance by fitting the data:
\begin{equation}
\begin{split}
\log{\tilde{\tau}_{\rm g}} = - 1.99 \times |b|^{0.156} + 0.321 \times d \times H(8.3-d) + 2.66 \times H(d-8.3) - 3.79 \,.
\end{split}
\label{eq:tau_Galaxy}
\end{equation}
The fluctuation scattering intensity can be obtained through difference calculations: $\log{\tilde{\tau}_{\rm s}} = \log {\tilde{\tau}} - \log{\tilde{\tau}_{\rm g}}$.
$\log{\tilde{\tau}_{\rm s}}$ allows us to exclude the effects of distance and galactic latitude, revealing more about the influence of specific structures along the line of sight on scattering (see the right panel of Fig.~\ref{fig:tau_tautl_tautls}). %This also indirectly indicates fluctuations in electron number density along these sightlines, aiding in our precise understanding of the structure of the interstellar medium. 

With the fluctuation scattering intensity, one can study statistically how spatially correlated pulsar scattering is. Additionally, coherent patterns can be searched to identify foreground scattering structures, as will be presented in Section.~\ref{Results}.
%, which we will show in the following. One can also search for coherent patterns to identify foreground scattering structures, which we shall present in Section.~\ref{Results}.

%At the moment, G38 remains the only structure discovered using the distribution of $\tilde{\tau}_{\rm s}$. However, with the increase of pulsar scattering data, we expect the fitting formula (\ref{eq:tau_Galaxy}) will aid the discovery of more coherent patterns in pulsar scattering and structures in the ISM. %Fig.~\ref{fig:tau_tautl_tautls} shows the Galactic distribution of pulsars in the sample with $|z| < 0.4$\,kpc. \xun{In the three panels, the color of the data points indicates the values of the pulsar scattering time ${\tau}$, the reduced pulsar scattering strength $\tilde{\tau}$, and the fluctuation scattering intensity $\tilde{\tau}_s$, respectively.}

\subsubsection{Correlation coefficient}
To constrain the typical size of pulsar scattering screens, we study the spatial correlation of the fluctuation scattering intensity $\tilde{\tau}_{\rm s}$.
We group pulsar pairs with different separations up to 1.5\,kpc into 10 evenly spaced bins. For the $i$th group, the correlation coefficient $R_{i}$ of the pulsars is given by
\begin{equation}
{R_{i}} = \frac{\left [ \sum_{j} \left ( \tilde{\tau}_{{\rm A}_{ij}} - {\langle \tilde{\tau} \rangle}_{i} \right )\cdot  \left ( \tilde{\tau}_{{\rm B}_{ij}} - {\langle \tilde{\tau} \rangle}_{i} \right ) \right ]  / {\rm N}_{i}}{\left \{  \sum_{j} \left [ \left (  \tilde{\tau}_{{\rm A}_{ij}} - {\langle \tilde{\tau} \rangle}_{i} \right ) ^{2} + \left (  \tilde{\tau}_{{\rm B}_{ij}} - {\langle \tilde{\tau} \rangle}_{i} \right ) ^{2} \right ] \right \} / 2{\rm N}_{i}} 
\label{eq:Diffi}
\end{equation}
where $\langle \tilde{\tau} \rangle_{i}$ represents the average $\tilde{\tau}_{\rm s}$ of the $i$th group of pulsars, $\tilde{\tau}_{{\rm A}_{ij}}$ and $\tilde{\tau}_{{\rm B}_{ij}}$ are the two $\tilde{\tau}_{\rm s}$ values of the $j$th pair of pulsars, and ${\rm N}_{i}$ denotes the number of pulsar pairs in the $i$th group. To evaluate the uncertainty in the mean correlation coefficient for each group, we use the median absolute deviation (MAD) to calculate the error bars. Unlike standard deviation, MAD is less sensitive to outliers and provides a more robust measure of dispersion.

\begin{figure*}
    \centering
	\includegraphics[width=10cm]{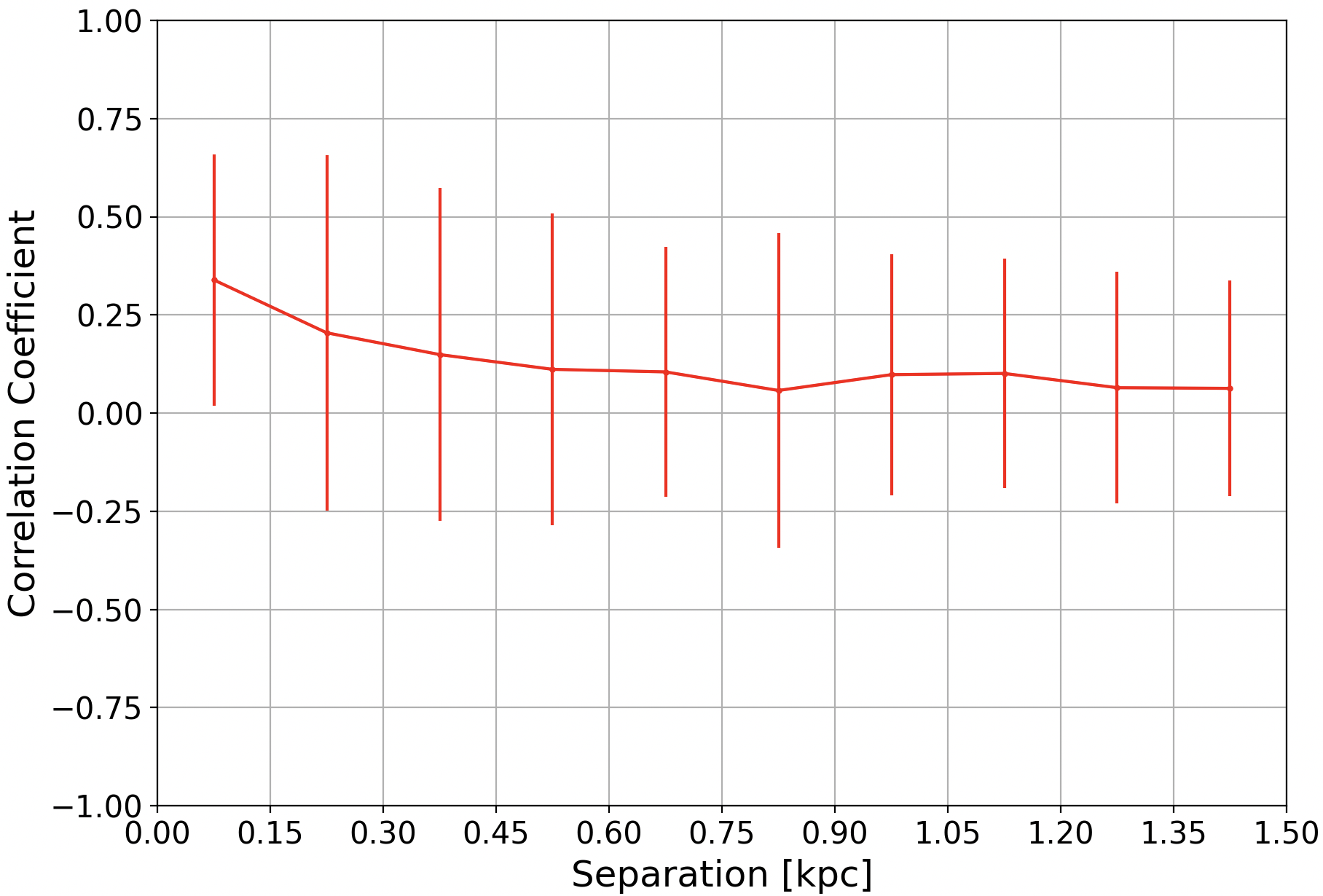} 
    \caption{\textbf{The correlation coefficient of fluctuation scattering intensity $\tilde{\tau}_{\rm s}$}, with the distance between pairs as the x-axis and the degree of the correlation coefficient between pulsar pairs as the y-axis. The red line represents the average correlation coefficient value of scattering strength for each distance group, along with the error values.}
    \label{fig:ACF}
\end{figure*}

Fig.\ref{fig:ACF} presents the correlation coefficient for the fluctuation scattering intensity $\tilde{\tau}_{\rm s}$ of pulsar pairs. A slight positive correlation is observed at separations smaller than 0.45\,kpc, with the highest correlation occurring in the first bin (median separation of $\sim$0.1\,kpc), where the correlation is still weak, with a coefficient of about 0.3. 
%Although the correlation observed in the first bin is weak, this may be due to the limited number of pulsar pairs at small separations. 
The current grouping with a bin size of 0.15\,kpc already represents the practical limit given the sparsity of close pulsar pairs. 
%However, the trend in Fig.\ref{fig:ACF} suggests the possibility of stronger correlations at separations smaller than 0.1\,kpc.} 
The current density of pulsars with scattering data is not yet sufficient to resolve the typical spatial scale of pulsar scattering screens in the ISM. Nevertheless, our result rules out the dominance of pulsar scattering by structures larger than 0.15\,kpc.

%\hqy{Previous studies also support the idea that small-scale structures dominate pulsar scattering. \citet{B1133} report that a structure as small as $\sim$ 0.0035\,pc could cause scattering for the pulsar B1133+16 at a distance of 186\,pc, and the Sh 2-27 structure, with a size of 34\,pc \citep{Sh227}, is capable of scattering the pulsar PSR J1643-1224 \citep{J16431, J16432}, we also identified structure G38 with a size of 50\,pc, also falls well below the 0.1\,kpc scale. Future observations with more densely sampled pulsar scattering data are needed to %verify whether the correlation increases at separations smaller than 0.1\,kpc and to conclusively establish the dominant scale of the scattering structures.}

\section{Galactic Structures Affecting Pulsar Scattering}
\label{Results}

Identifying and characterizing structures that influence pulsar signal scattering is crucial for understanding the connection between pulsar scattering and the spatial distribution of the ISM. In this section, 
we use the Galactic distribution of pulsar fluctuation scattering intensity $\tilde{\tau}_{\rm s}$ to uncover spatially coherent patterns, and correlate them with other ISM observation data to identify ISM structures that may significantly affect scattering.
We identify two structures associated with prominent spatial patterns in our pulsar data (Table~\ref{tab:list}): the Vela supernova remnant within the Gum Nebula, whose influence on pulsar scattering data is well known, and G38, a distant superbubble structure newly discovered in this paper.

\begin{table*}[h]
\centering
\caption{A list of Galactic structures affecting pulsar scattering. Their positions in the Galaxy can be seen in Fig.~\ref{fig:Galaxy}. The symbol $\dagger$ indicates that the value is derived from the results presented in \citet{Cordes02}.}
\begin{tabular}{|c|c|cc|cc|c|}
\hline
\multirow{2}{*}{NAME} & Distance & \multicolumn{2}{c|}{Size (Diameter)} & \multicolumn{2}{c|}{Range of impact on pulsar scattering} & \multirow{2}{*}{references}\\ \cline{3-6}
 & kpc    & \multicolumn{1}{c|}{kpc}   & degree  & \multicolumn{1}{c|}{l}     & b  &      \\ \hline
\multirow{5}{*}{Vela supernova remnant}           & \multirow{5}{*}{0.25$^{\dagger}$}    & \multicolumn{1}{c|}{\multirow{5}{*}{0.08$^{\dagger}$}}           & \multirow{5}{*}{$\sim$18$^{\circ}$$^{\dagger}$}  & \multicolumn{1}{c|}{\multirow{5}{*}{254$^{\circ} \sim$271$^{\circ}$}}           & \multirow{5}{*}{-15$^{\circ} \sim$-2$^{\circ}$}     &\citet{Vela0, Vela1}  \\ 
 & & \multicolumn{1}{c|}{} & & \multicolumn{1}{c|}{} & & \citet{Vela2} \\
 & & \multicolumn{1}{c|}{} & & \multicolumn{1}{c|}{} & & \citet{Vela3} \\
 & & \multicolumn{1}{c|}{} & & \multicolumn{1}{c|}{} & & \citet{Vela4} \\
  & & \multicolumn{1}{c|}{} & & \multicolumn{1}{c|}{} & & \citet{Cordes02, Cordes03}... \\ \hline
G38               & 2.3     & \multicolumn{1}{c|}{0.05}         & $\sim$1$^{\circ}$ & \multicolumn{1}{c|}{37$^{\circ} \sim$ 38.5$^{\circ}$}              & 0.5$^{\circ} \sim$ 2.5$^{\circ}$       & \textbf{This paper}  \\ \hline   
\end{tabular}
\label{tab:list}
\end{table*}

\subsection{Other observation data of the ISM}

To validate the correlation between pulsar scattering data and ISM structures and explore the intricate relationship between pulsar scattering and the Galactic environment, we compare our pulsar scattering data with the following survey data in our study:

(a) Dust Extinction Map: \citet{dustmap} combined Gaia DR2 photometric data with Two Micron All Sky Survey measurements to create a three-dimensional map of interstellar dust in the Local Arm and surrounding regions.

(b) H$_{\alpha}$ Full Sky Map: By combining the Virginia Tech Spectral Line Survey \citep{VTSS} from the north and the Southern H$_{\alpha}$ Sky Survey Atlas \citep{SHASSA} from the south, \citet{Halpha} created the H$_{\alpha}$ Full Sky Map.\footnote{\url{https://faun.rc.fas.harvard.edu/dfink/skymaps/halpha/}}
The Wisconsin H$_{\alpha}$ Mapper survey \citep{WHAM} covered three-quarters of the sky at a 1$^{\circ}$ scale, providing a stable zero-point reference.

(c) RRLs: Radio recombination lines are primarily observed in ${\rm H}\,_{\rm II}$ regions formed by radiation from massive stars. They provide vital information about the electron density, temperature, and ionization structure of the ISM. In this paper, we correlate the sensitive average $\rm H_{n\alpha } $ RRL map covering 88 deg$^2$ in the inner Galaxy of $33^{\circ } \le l\le 55^{\circ }$ and $\left | b \right | \le 2.0^{\circ } $ by \citet{Hou2} constructed using the FAST Galactic Plane Pulsar Snapshot survey (GPPS) data taken with the Five-hundred-meter Aperture Spherical radio Telescope 
\citep[FAST;\footnote{\url{http://zmtt.bao.ac.cn/GPPS/}}][]{GPPS}.

(d) Planck Map: Its 857\,GHz channel is capable of detecting microwave radiation emitted by dust, providing structural information about dust and gas in the Galaxy \citep{Plank19}.

(e) HI4PI: The \citet{HI4PI} integrated the Effelsberg–Bonn ${\rm H}\,_{\rm I}$ Survey \citep{HI16} and the Galactic All-Sky Survey \citep{HI10} data into a comprehensive and unified dataset of neutral hydrogen observations, known as the ${\rm H}\,_{\rm I}$ 4$\pi$ survey.\footnote{\url{http://cdsarc.u-strasbg.fr/viz-bin/qcat?J/A+A/594/A116}}
This dataset provides velocity information that lets us obtain the kinetic distances \citep{v_d16, v_d19} for structures of our interest.\footnote{\url{http://bessel.vlbi-astrometry.org/node/378}}

(f) WISE: Using data from the Wide-field Infrared Survey Explorer, \citet{WISE} identified over 8000 Galactic ${\rm H}\,_{\rm II}$ regions and candidates, creating the most comprehensive catalog of massive star formation areas.\footnote{\url{http://astro.phys.wvu.edu/wise}}

\subsection{Vela Supernova Remnant}

\begin{figure}
    \centering
	\includegraphics[width=10cm]{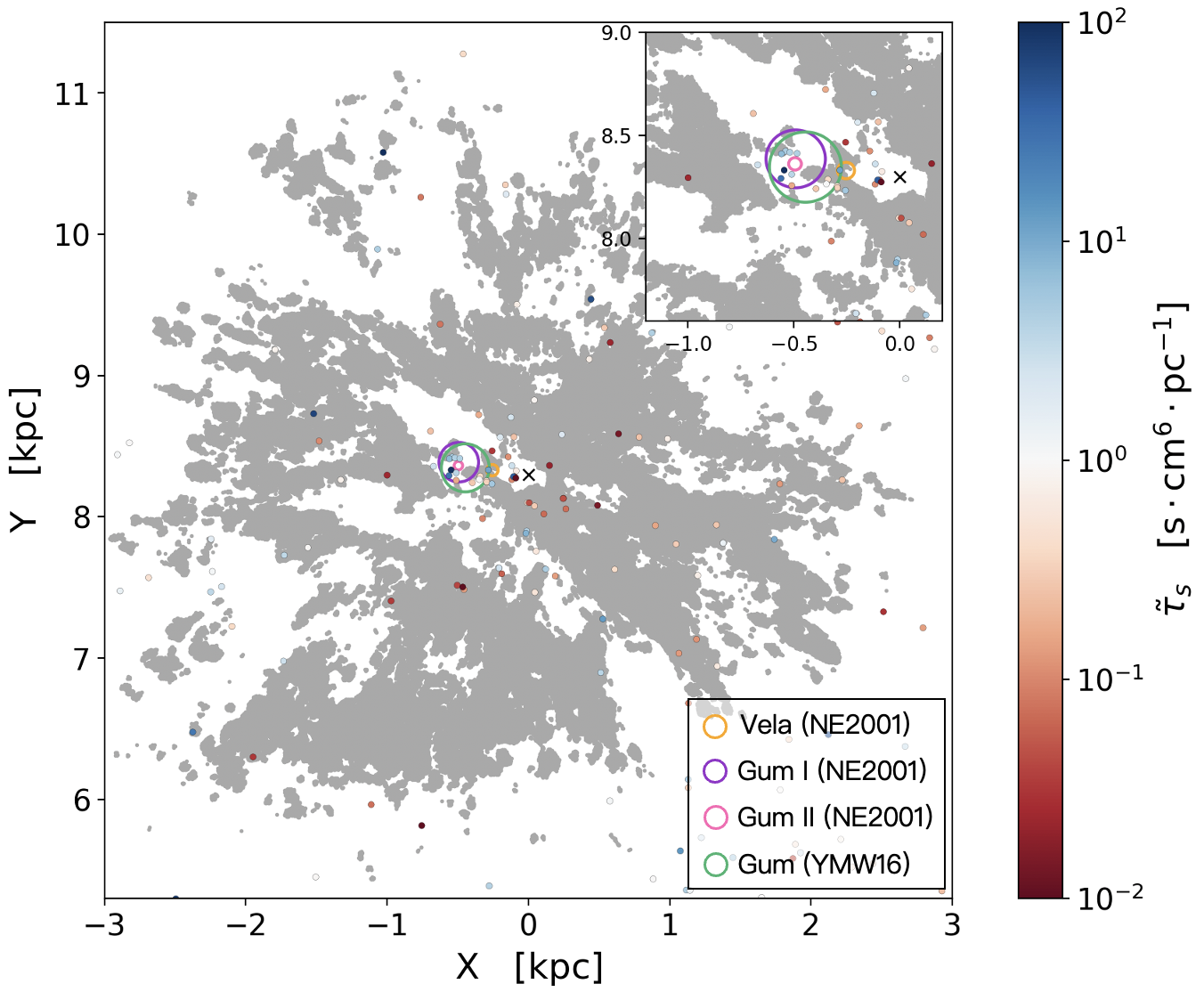}\\
    \includegraphics[width=10cm]{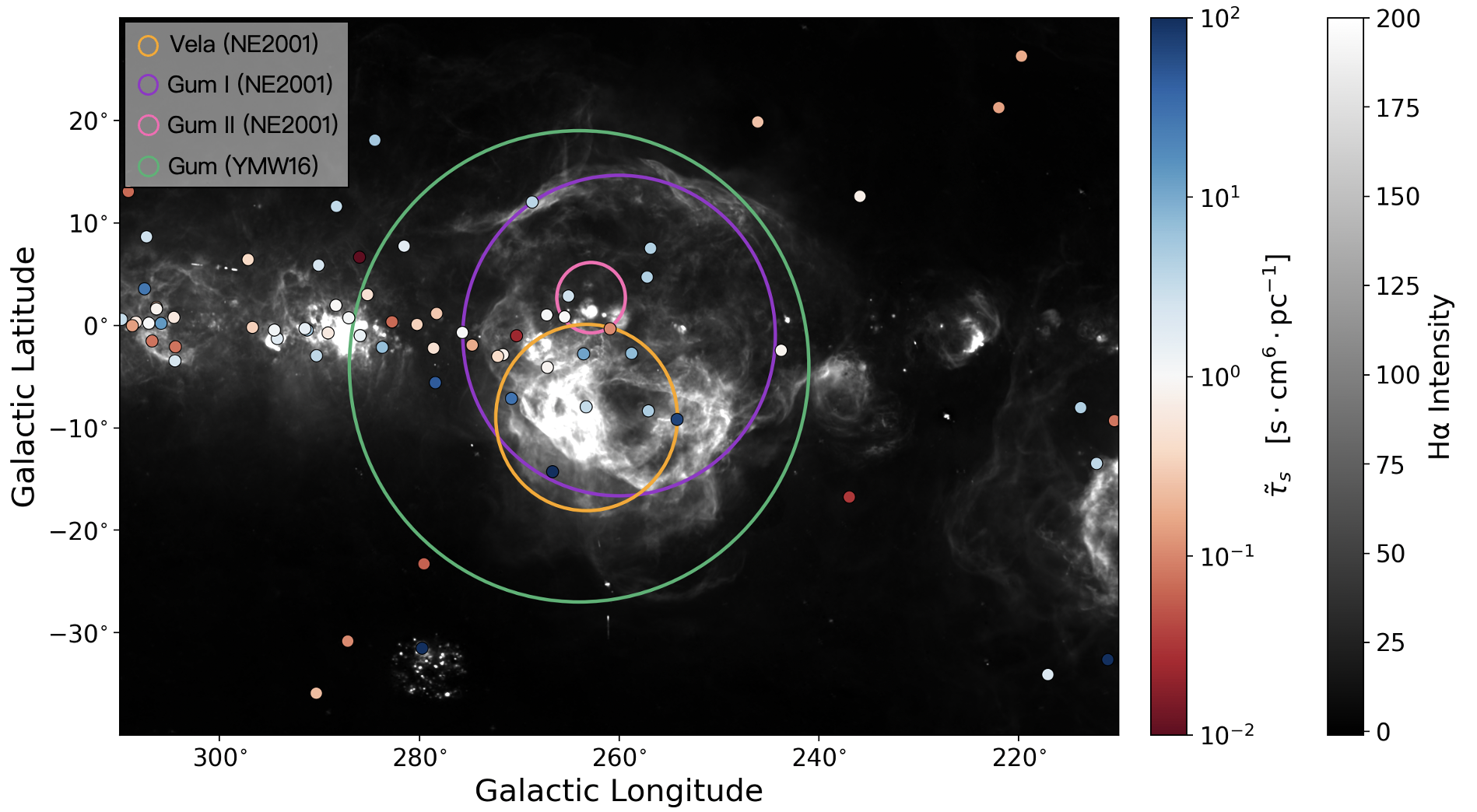}
    \caption{\textbf{Gum Nebula/Vela affects pulsar fluctuation scattering intensity.} 
    The top is the fluctuation scattering intensity ($\tilde{\tau}_{\rm s}$) distribution of local pulsars with the dust extinction map as the background. 
    %The pink and orange circles respectively indicate the location of the Vela and Gum Nebula as given by the ``NE2001'' model \citet{Cordes02, Cordes03} \hqy{(where the orange circles represent Gum\,I (larger circle) and Gum\,II (smaller circle), with Gum\,I having higher electron density)}, and t
    The green circle represents the Gum Nebula location according to the YMW16 model \citet{YMW16}, whereas the orange, purple, and pink circles indicate the three components in the Gum region identified in the NE2001 model \citet{Cordes02, Cordes03} -- the Vela Nebula, Gum\,I, and Gum\,II, respectively.
    The inset provides a magnified view of the Gum Nebula/Vela region for enhanced detail.
    The bottom is the fluctuation scattering intensity ($\tilde{\tau}_{\rm s}$) distribution of pulsars near the $l$-$b$ range of the Gum Nebula/Vela, with the H$_{\alpha}$ map as the background. The circles in the figure represent the same meanings as those in the top panel.} 
    %(a cone region in the 3D distribution, as shown in the inset).}
    \label{fig:dust}
\end{figure}

The local population of pulsars shows a distinct coherent pattern in their reduced scattering intensity (see the middle of Fig.~\ref{fig:tau_tautl_tautls}). Here we adopt the Galactic center coordinate system, using the Sun’s position at (0, 8.3)\,kpc as in the YMW16 model by \citet{YMW16}. We map the local distribution of fluctuation scattering intensity ($\tilde{\tau}_{\rm s}$) and examine its correlation with the distribution of dust map, as shown at the top of Fig.~\ref{fig:dust}.
%we present the local distribution of the reduced scattering intensity and its correlation with the local dust distinction map in Fig.~\ref{fig:dust}, where we have adopted a coordinate system centered on the Sun.
%So we map the reduced scattering intensity of the local pulsar samples within the range of $10^{-8.1} < \tilde{\tau }< 10^{-5.1}$ s\, cm$^{6}$\,pc$^{-1}$ and perform a correlation analysis with the dust extinction map. Given that dust maps are based on a 2D plane coordinate system centered on the Sun, which significantly differs from the galactic center coordinate system, we convert the dust map coordinates to galactic center coordinates (where, $X = Y_{\rm dust}/1000$; $Y = -(X_{\rm dust}/1000 - 8.3)$), and then selected pulsar samples within the coverage range of the dust maps (-3\,kpc $\le X \le $ 3\,kpc, 5.3\,kpc $\le Y \le $ 11.3\,kpc) for analysis. Fig.~\ref{fig:dust} illustrate the correlation results of our pulsar samples with the dust map in X-Y plane and 3D space. It is observed that pulsars located within the dust have weaker scattering compared to those on the periphery, which we believe may be due to the lines of sight of the peripheral pulsars passing through more structural boundaries, leading to increased scattering. 
It is noteworthy that there exists an elliptical dust void centered around $x = -0.5$\,kpc, $y = 8.6$\,kpc. This void approximately spans around -1\,kpc $\le X \le $ 0.1\,kpc and 8.2\,kpc $\le Y \le $ 8.9\,kpc. 
%(where, $X = Y_{\rm dust}/1000$; $Y = -(X_{\rm dust}/1000 - 8.3)$, \hqy{$X_{\rm dust}$ and $Y_{\rm dust}$ represent the $X$ and $Y$ values in the dust data}). 
This elliptical dust void corresponds to the Local Superbubble \citep{LB, Cordes02}, whose formation is driven by multiple supernova explosions and powerful stellar winds. 
The Gum Nebula \citep{Gum52}, located in the negative Y-axis portion of this area and spatially overlapping with the Local Superbubble, is believed to significantly alter the scatter of pulsars located within or behind it. The Gum Nebula is a complex region that contains multiple star clusters, young stars, and other nebulae with high temperatures \citep{Gumtem2, Gumtem3, Gumtem1, Gumtem4, Gumtem5}. 
Within this region, the Vela supernova remnant, a single coherent structure formed by the explosion of the Vela supernova \citep{Vela}, is more compact and relatively closer to Earth \citep{Vela0, Vela1, Vela2, Vela3, Vela4}.
The same explosion generated the famous Vela pulsar, whose scintillation arcs were recently detected by \citet{Xu_Vela}, revealing a scattering screen at 0.19$\pm$0.03\,kpc from Earth that likely corresponds to the boundary of the Local Superbubble.
%\xun{//what does `its scattering effect being equally remarkable' mean here? By the way, yonghua xu first detected a scintillation arc towards the Vela pulsar. maybe we can cite his paper too}

The impact of these structures has been incorporated into Galactic electron distribution models, and extensively mentioned in Galaxy models, especially regarding the discussion of these structures' impact on the local electron density. Due to the lack of sufficient pulsar scattering measurements in the early years along these sightlines, the TC93 model \citep{Taylor93} did not consider the Gum Nebula region to affect scattering. However, the NE2001 model \citep{Cordes02, Cordes03} incorporated scattering measurements from five pulsars to define the Gum Nebula/Vela area (indicated by orange, purple, and pink circles in Fig.~\ref{fig:dust}). Similarly, the YMW16 model \citep{YMW16} used DMs from eight pulsars to delineate the Gum Nebula region (indicated by the green circle in Fig.~\ref{fig:dust}). 
%This progressive refinement highlights the evolving understanding of interstellar scattering effects in complex regions of the Galaxy. 

We conduct a comparative analysis of pulsar fluctuation scattering intensity ($\tilde{\tau}_{\rm s}$) with H$_{\alpha}$ map for the Gum Nebula/Vela regions within the $l$-$b$ range (see the bottom of Fig.~\ref{fig:dust}). Compared to the Gum Nebula, the Vela region exhibits significantly more pronounced pulsar scattering characteristics. The Gum Nebula covers a larger spatial area, and its impact on pulsar scattering is comparable to the influence caused by other structural features in the ISM within this region. In contrast, the Vela's more notable scattering effects suggest that it possesses higher electron density fluctuations. According to the scattering data, the Vela region's influence on pulsar scattering spans approximately $17^{\circ}$ in the Galactic longitude direction, centered at $l = 262.5^{\circ}$, and about $13^{\circ}$ in the Galactic latitude direction, centered at $b = -8.5^{\circ}$.

This underscores the potential of pulsar scattering data to reveal different aspects of ISM structures -- 
it may be challenging to distinguish between the Vela and Gum regions using direct observations from surveys such as HI4PI, WISE, and Planck.
%might otherwise be difficult to directly observe through surveys like HI4PI, WISE, and Planck. 

\subsection{G38: A distant superbubble}
\label{G38}

\begin{figure*}
    \includegraphics[width=\linewidth]{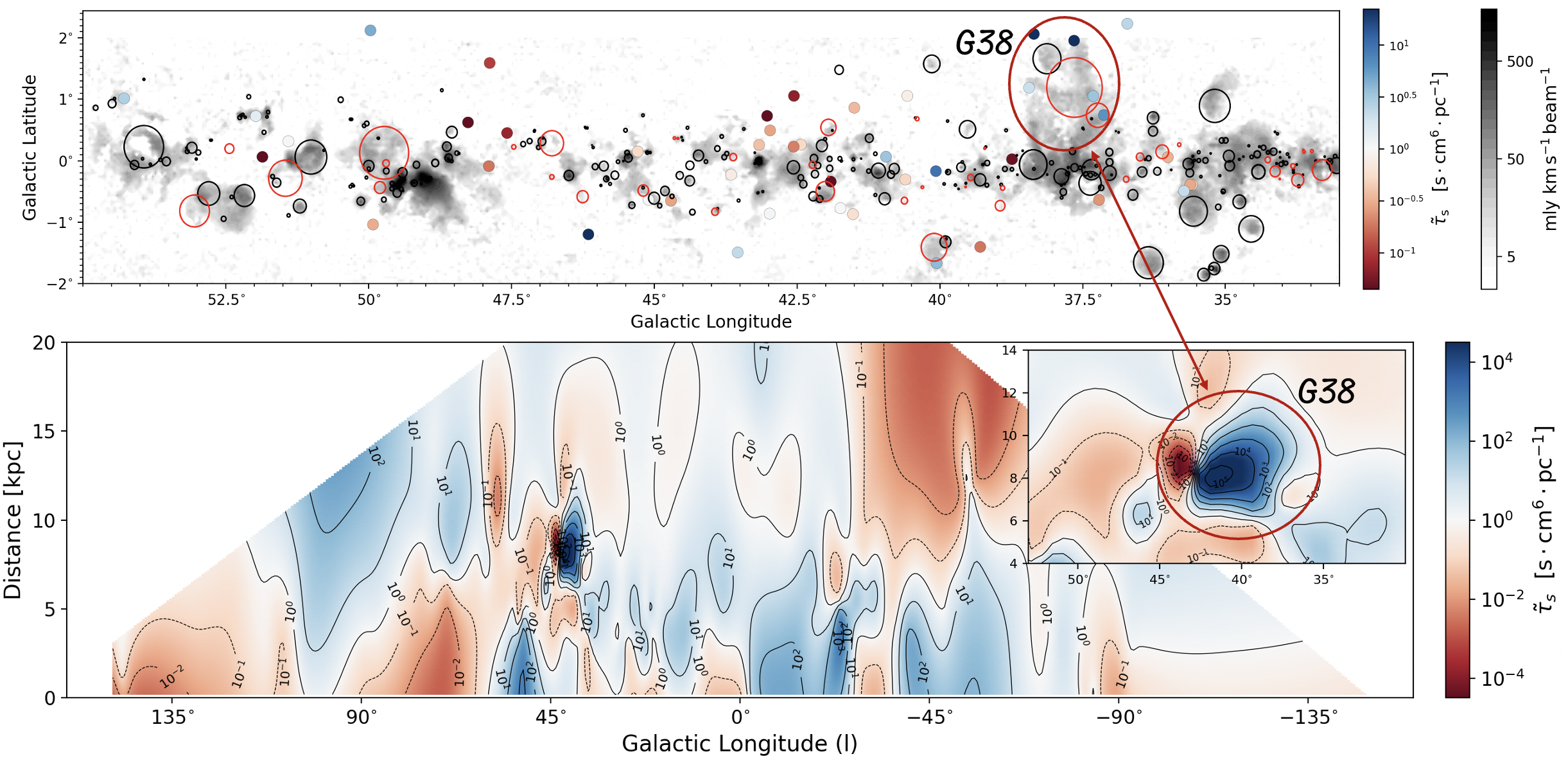}
    \caption{\textbf{G38 affects pulsar fluctuation reduced scattering intensity.} The top is the distribution of fluctuation reduced scattering intensity ($\tilde{\tau}_{\rm s}$) in a patch of the sky (covering $33^{\circ} < l < 55^{\circ}$ and $-2^{\circ} < b < 2^{\circ}$) with \citet{Hou2} FAST RRL observations as a background. Newly certified ${\rm H}\,_{\rm II}$ regions are marked with red circles, while previously recognized ${\rm H}\,_{\rm II}$ regions are indicated by black circles. The bottom is the interpolated distribution of fluctuation scattering intensity ($\tilde{\tau}_{\rm s}$) of Galactic plane pulsars with $0.5^{\circ} \le b \le 2.5^{\circ}$ as a function of the Galactic longitude and distance. 
    The G38 region stands out as a coherent region much more heavily scattered than its surroundings.}
    %The \hqy{deeper blue} indicate the existence of a coherent foreground structure.}
    \label{fig:RRL}
\end{figure*}

In our pulsar scattering data, we identify a spatially coherent group of five pulsars 
%(J1853$+$0505, J1853$+$0545, J1855$+$0422, J1856$+$0404, and J1857$+$0526) 
with higher fluctuation reduced scattering intensity than their surroundings (see Table~\ref{tab:RRL_data}). To further examine whether the pulsar reduced scattering intensity in this region of the Galaxy is significantly higher than other pulsars at the same latitude, we interpolate the fluctuation reduced scattering intensity for pulsars within $0.5^{\circ} \le b \le 2.5^{\circ}$. A uniform grid is constructed based on Galactic longitude ($l$) and distance ($d$), and the grid-based interpolation method \texttt{gdatav4} is used to estimate the reduced scattering intensity across the grid. To ensure the focus remained on relevant scattering regions, we exclude areas in the anti-Galactic center direction with large distances that lacked sufficient pulsar scattering data. The resulting interpolated distribution (see bottom of Fig.~\ref{fig:RRL}) clearly illustrates the spatial variation in the fluctuation reduced scattering intensity in this Galactic region inferred from pulsar scattering data.
%To further examine whether the pulsar scattering intensity in this part of the Galaxy is significantly higher than other pulsars at the same latitude, we interpolate the reduced scattering intensity for pulsars within $0.5^{\circ} \le b \le 2.5^{\circ}$.

\begin{table*}[h]
	\centering
	\caption{The list of five pulsars influenced by the newly identified Distant Superbubble G38, Where the pulsar \textit{l}, \textit{b}, \textit{d}, and DM data Are from the ATNF catalog \citep{ATNF}.}
    \label{tab:RRL_data}
	\begin{tabular}{|c|ccccccc|} 
		\hline
\multirow{2}{*}{JNAME}      & \textit{l}      & \textit{b}    & \textit{d}      & DM  & $\log{\tilde{\tau }} $ & $\log{\tilde{\tau }_{\rm s}} $  & \multirow{2}{*}{Reference} \\ 
 & ($^{\circ}$) & ($^{\circ}$) & (kpc) & (pc$\cdot$cm$^{-3}$ )  & (s$\cdot$cm$^{6}\cdot$pc$^{-1}$) &  (s$\cdot$cm$^{6}\cdot$pc$^{-1}$) &  \\ \hline
J1853+0505 & 37.65  & 1.956 & 9.127  & 279.0 & -1.040& 2.299 & \citet{17.2004ApJ...605..759B} \\
J1853+0545 & 38.354 & 2.064 & 6.546  & 198.7 & -2.367&1.550 & \citet{58.2021MNRAS.504.1115O, 17.2004ApJ...605..759B} \\
J1855+0422 & 37.314 & 1.052 & 10.948 & 455.6 & -2.566& 0.569 & \citet{17.2004ApJ...605..759B} \\
J1856+0404 & 37.128 & 0.745 & 6.213  & 341.3 & -2.917& 0.779 & \citet{17.2004ApJ...605..759B} \\
J1857+0526 & 38.438 & 1.187 & 12.257 & 466.4 & -2.819&0.355 & \citet{58.2021MNRAS.504.1115O, 17.2004ApJ...605..759B} \\
    \hline
	\end{tabular}
\end{table*}

To confirm the existence of the structure, we compared other data and found a newly certified ${\rm H}\,_{\rm II}$ region in the $33^{\circ} \le l \le 55^{\circ}$ region given by \citet{Hou2}. For our sample of pulsars in the same sky region, we selected pulsars that lie beyond their nearest ${\rm H}\,_{\rm II}$ regions based on distance measurements, and correlated them
%we filter the pulsars \xun{//what do you mean by `filter' here?} with ${\rm H}\,_{\rm II}$ regions as foregrounds based on the distance information of the ${\rm H}\,_{\rm II}$ regions and correlated 
with the RRL data (see Fig.~\ref{fig:RRL}, the colors of the pulsars represent their fluctuation scattering intensity $\tilde{\tau}_{\rm s}$%, which will be presented in Section~\ref{Scattering_fits}
). We use the RRL data as background, and the red and black circles indicate the newly certified and known ${\rm H}\,_{\rm II}$ regions on \citet{Hou2} and WISE data, respectively. We find that our sample of those five pulsars is behind one of their newly certified ${\rm H}\,_{\rm II}$ region structures, G37.643$+$1.193, which has a sufficiently large radius of 1747'' and a distance of 2.3 $\pm$ 0.4\,kpc \citep{Hou2}. At the same time, we also see a similar structure in the HI4PI data at a velocity of around 43.9 km/s.
Using the Galactic dynamics model \citep{v_d14, v_d19}, this gives an approximate distance of 2.39$\pm$0.35\,kpc for this structure, in agreement with the RRL data results.

\begin{figure}[h]
    \centering
	\includegraphics[width=10cm]{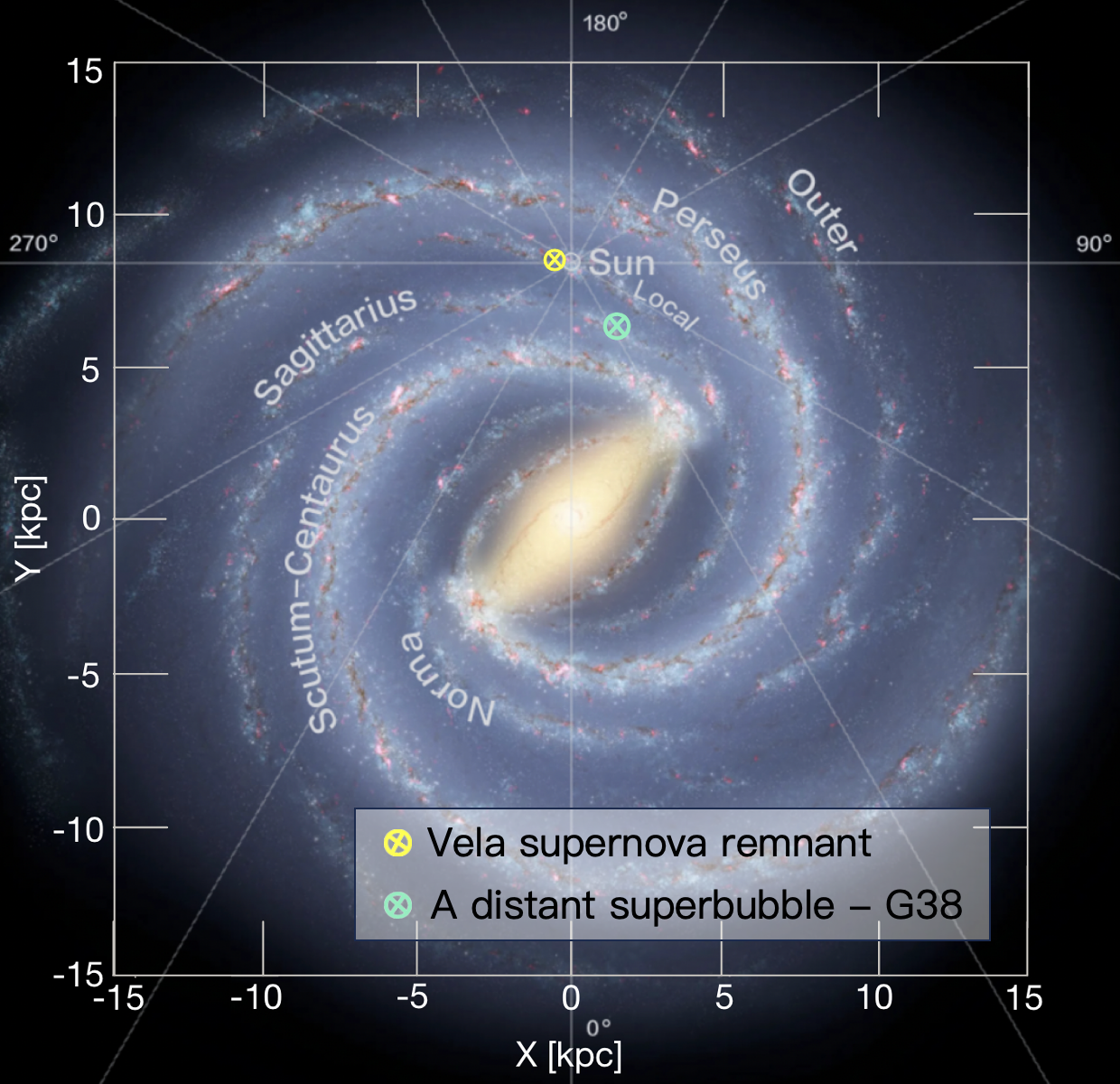}
    \caption{Locations of the Vela supernova remnant and G38 in our Galaxy. The background image is a schematic of the Galactic disk as viewed from the Northern Galactic Pole (courtesy of NASA/JPL-Caltech/R. Hurt (SSC/Caltech)).}% The circles represent their locations within the Milky Way and do not indicate their sizes.}
    \label{fig:Galaxy}
\end{figure}

It is noteworthy that an established ${\rm H}\,_{\rm II}$ region, G38.124$+$1.661 (marked by a black circle), is located near G37.643$+$1.193. The radial velocity ($V_{\rm LSR}$) of G38.124$+$1.661 is 41.7 $\pm$ 0.4\,km\,s$^{-1}$ \citep{WISE}, and its kinematic distance, derived from the Galactic rotation model \citep{v_d14, v_d19}, is %calculated to be 
2.32 $\pm$ 0.32\,kpc. This distance is comparable to that of G37.643$+$1.193. 
%These two ${\rm H}\,_{\rm II}$ regions likely have contributed to the formation of a superbubble, which in turn affects the scattering cross-section of background pulsars in this area. 
At a distance scale of 2.3\,kpc, these two ${\rm H}\,_{\rm II}$ regions are located near the Sagittarius Arm and have collectively formed a superbubble structure (G38), significantly impacting the scattering cross-section of background pulsars in this region. We take the center-to-center distance between G38.124$+$1.661 and G37.643$+$1.193 as the angular radius of the distant superbubble G38. Scaling it by its distance of 2.3\,kpc, the physical size of the bubble is estimated to be approximately 50\,pc. Fig.~\ref{fig:Galaxy} shows more visually the position of the Vela supernova remnant and G38 in our Galaxy.
%The background image is a schematic of the Galactic disk as viewed from the Northern Galactic Pole (courtesy of NASA/JPL-Caltech/R. Hurt (SSC/Caltech)).

%The scale of this distant superbubble G38 is estimated to be 50\,pc based on the center-to-center distance between G38.124$+$1.661 and G37.643$+$1.193.

\subsection{Correlation with other data and discussion}
\label{se:correlation}

In addition to dust extinction and RRL data, we also correlate our pulsar scattering sample with Planck 857\,GHz \citep{Plank19}, HI4PI \citep{HI4PI}, and the ${\rm H}\,_{\rm II}$ region catalog from the WISE data \citep{WISE}. The results for correlation with Planck data and the ${\rm H}\,_{\rm II}$ catalog from WISE data are presented in Appendix~\ref{ap:plankwise}. 
No correlated pattern is detected with the Planck 857\,GHz map.
%We find the overlap between known ${\rm H}\,_{\rm II}$ regions and pulsars with scattering data is very small, implying that the impact of these ${\rm H}\,_{\rm II}$ regions are limited. No correlated pattern has been detected with the Planck 857\,GHz map.

Although we find little overlap between known ${\rm H}\,_{\rm II}$ regions and pulsar scattering data, this does not imply that the influence of ${\rm H}\,_{\rm II}$ regions on pulsar scattering can be disregarded. \citet{Ocker24} confirmed the significant impact of known ${\rm H}\,_{\rm II}$ regions as scattering media on the signals of pulsars behind them by conducting a detailed analysis of the spatial relationships between a large number of pulsars and ${\rm H}\,_{\rm II}$ regions, substantiating the pervasive nature of ${\rm H}\,_{\rm II}$ regions' influence on pulsar scattering. However, in their study, they employed a twofold radius for the ${\rm H}\,_{\rm II}$ regions' area of influence, discovering that 630 pulsars were affected, while only 220 pulsars were within a onefold radius. They obtained the potential DM contribution from ${\rm H}\,_{\rm II}$ regions but did not explicitly quantify the difference in scattering parameters between pulsars inside and outside ${\rm H}\,_{\rm II}$ regions.

In contrast, we attempt to directly utilize pulsar scattering data to investigate ISM structures at various scales that may influence scattering intensity, including ${\rm H}\,_{\rm II}$ regions and other patterns, facilitating the discovery of new scattering structures. However, the identification of fewer overlaps in our study can be attributed to both the current sparsity of pulsar scattering measurements and our methodological requirement of multiple pulsars exhibiting enhanced scattering within the same region to confirm a new structure. Nevertheless, our approach demonstrates the viability and potential of pulsar scattering measurements as a probe of ISM structures.

Considering both \citet{Ocker24}'s results and our analysis, we suggest that as high-quality pulsar scattering measurements increase in the future, the number of pulsars intersecting ${\rm H}\,_{\rm II}$ regions will likely exceed our current findings but remain below \citet{Ocker24}'s estimates.

%\hqy{In Fig.~\ref{fig:all_data}, we show our pulsar scattering data together with the Planck 857\,GHz map (upper panel) and verified ${\rm H}\,_{\rm II}$ regions in WISE (lower panel). }

%The results for correlation with Planck data and the ${\rm H}\,_{\rm II}$ catalog from WISE data are presented in Fig.~\ref{fig:all_data}. 
%We find the overlap between known ${\rm H}\,_{\rm II}$ regions and pulsars with scattering data is very small, making the impact of these ${\rm H}\,_{\rm II}$ regions very limited. No correlated pattern has been detected with the Planck 857\,GHz map.

\section{Conclusion}
We have explored the ability of pulsar scattering data in exploring ISM structures using a large dataset of 473 pulsar scattering time measurements.

By fitting pulsar scattering data, we have derived the smooth dependencies of the pulsar scattering intensity distribution on the Galactic latitude and distance: $\log{\tilde{\tau}_{\rm g}} = - 1.99 \times |b|^{0.156} + 0.321 \times d \times H(8.3-d) + 2.66 \times H(d-8.3) - 3.79$. 
%We believe that 
If large scattering structures exist, their imprint on the fluctuation scattering intensity, obtained by subtracting the smooth function, will be clearly distinguishable. 
However, our study of the correlation coefficient of the pulsar scattering distribution suggests that pulsar scattering is predominantly caused by structures smaller than 0.15\,kpc. To discover smaller scattering structures systematically would require more densely populated pulsar scattering data.

%We systematically explored ISM structures purely based on pulsar scattering data, while integrating a multiwavelength view of the ISM. 
Using the current dataset, we have discovered two prominent coherent structures on the distribution of the fluctuation scattering intensity. One is the Vela supernova remnant embedded in the Gum Nebula. The Gum Nebula is known to affect pulsar scattering. However, we have found that only the region covered by the Vela supernova remnant, which is a small fraction of the Gum Nebula, has significantly enhanced scattering compared to the surroundings. The other structure is a distant superbubble, G38 (37$^{\circ} < l <$ 38.5$^{\circ}$, 0.5$^{\circ} < b < $ 2.5$^{\circ}$), which we have identified using the pulsar scattering data. It has an approximate size of 50\,pc, %which significantly affects the scattering of pulsars located behind it. 
and is located at a distance of 2.3\,kpc in the Sagittarius Arm.
The discovery of G38 demonstrates the potential of pulsar scattering data in probing ISM structures. 

We have systematically compared the pulsar scattering distribution with other multiwavelength ISM data. The identities of the two structures we discovered using pulsar scattering data have both been confirmed by H$_{\alpha}$ data and RRL data, respectively. However, a general correlation between pulsar scattering and other ISM data is lacking. This is expected because pulsar scattering, by revealing the integrated electron density fluctuations along the line of sight, provides a complementary view of the ISM compared to the other probes.

In the future, by further enhancing the measurement and collection of pulsar scattering data, as well as leveraging more advanced telescopes and detection technologies, such as the upcoming Square Kilometre Array, a significant increase in both the quantity and the quality of pulsar scattering data is expected. With such future data more ISM structures can be detected with our method, and pulsar scattering can emerge as a new tool for ISM studies.

\begin{acknowledgments}
    \noindent We are grateful to the referee for valuable comments and constructive suggestions. This work is supported by NSFC grant No. 12373025. G.X.L. acknowledges support from NSFC grant Nos. 12273032 and 12033005.
\end{acknowledgments}

\appendix

\section{MCMC sampling for parameter fitting}
\label{ap:mcmc}
\renewcommand{\thefigure}{A\arabic{figure}}
\setcounter{figure}{0}

\begin{figure}
    \centering
	\includegraphics[width=10cm]{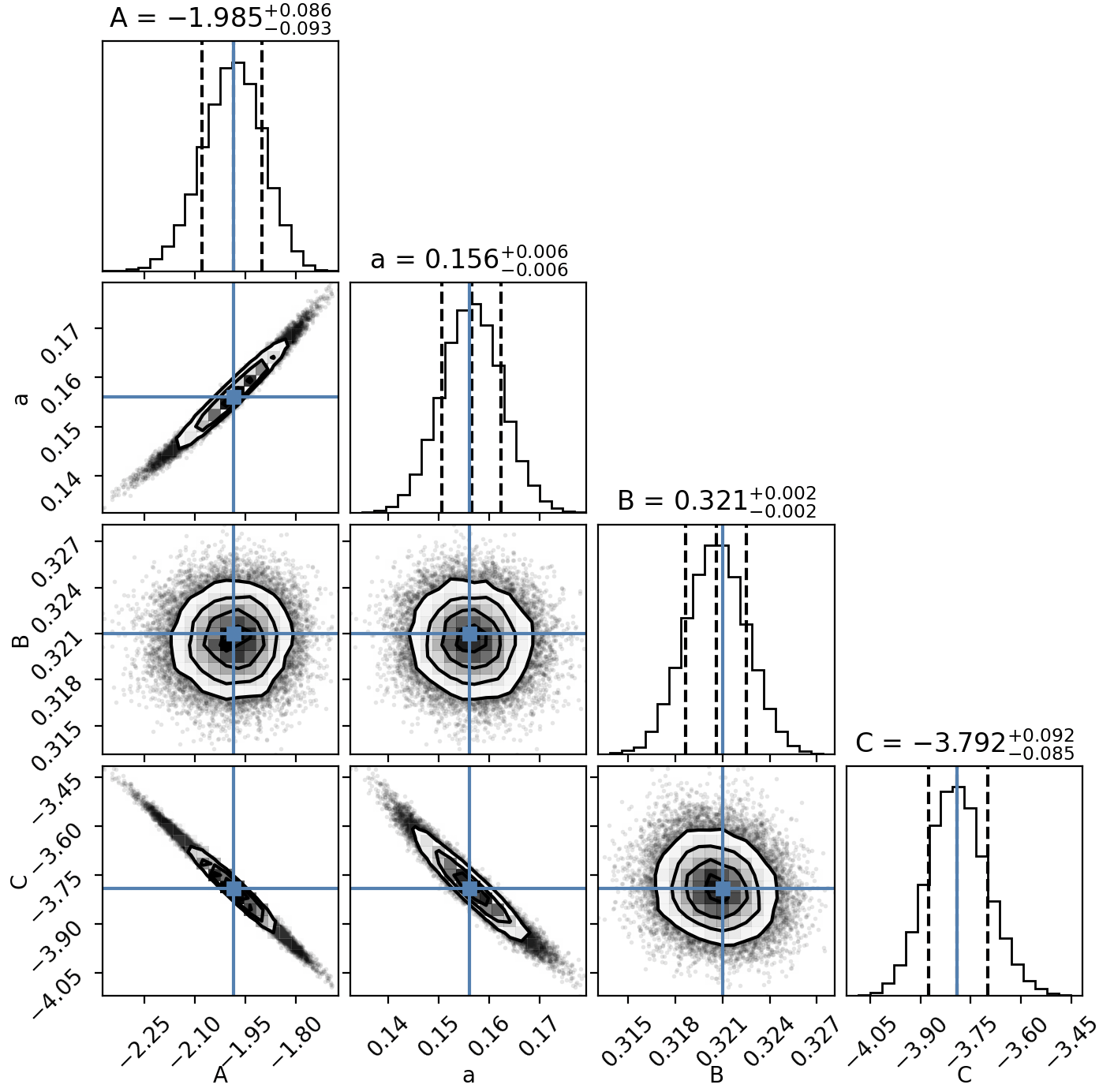}
    \caption{Posterior distributions and pairwise correlations of the model parameters described in Section~\ref{fitting}, based on the Equation~\ref{eq:fitting}.}
    \label{fig:mcmc}
\end{figure}

To robustly estimate the parameters \({\rm A}, {\rm a}, {\rm B}, {\rm C}\) in the reduced scattering intensity model, we employed the MCMC method. The MCMC sampling was performed using the Python package \texttt{emcee}, which efficiently explores the parameter space by combining a likelihood function and prior distributions to generate posterior distributions for the parameters. The sampling was initialized with 50 walkers near the least-squares solution, and after a burn-in phase of 100 steps, 5000 sampling steps were performed for each walker. The resulting posterior distributions of the parameters are shown in Fig.~\ref{fig:mcmc}. The diagonal panels show the marginalized posterior distributions for each parameter, while the off-diagonal panels display the pairwise correlations. The blue lines and markers indicate the median values of the parameters, with the shaded regions corresponding to the 1$\sigma$ uncertainties.

\section{Correlation with Planck and WISE Data}
\label{ap:plankwise}
\renewcommand{\thefigure}{B\arabic{figure}}
\setcounter{figure}{0}

Here, we show our pulsar scattering data together with the Planck 857\,GHz map (upper panel) and verified ${\rm H}\,_{\rm II}$ regions in WISE (lower panel).  Pulsars with stronger reduced scattering intensity (pulsars near the inner region) are concentrated on the galactic disk, whereas pulsars with weaker reduced scattering intensity (pulsars in the local region or at high Galactic latitudes) are widely distributed. No clear correlation has been detected in this direct comparison between pulsar scattering distribution and the Planck map. The lack of distance information in the Planck data prohibits a more detailed comparison.

In the correlation with WISE ${\rm H}\,_{\rm II}$ regions, we retain sources with pulsar distances greater than 4\,kpc, and limit ${\rm H}\,_{\rm II}$ regions to those closer than 4\,kpc, ensuring an adequate data sample size and that these ${\rm H}\,_{\rm II}$ structures are all within the foreground of our samples. However, the sample of ${\rm H}\,_{\rm II}$ regions with given distances, in particular those with large angular sizes, remains sparse. There exists barely any overlap between these ${\rm H}\,_{\rm II}$ regions and the pulsars with scattering data.

\begin{figure*}
	\includegraphics[width=\linewidth]{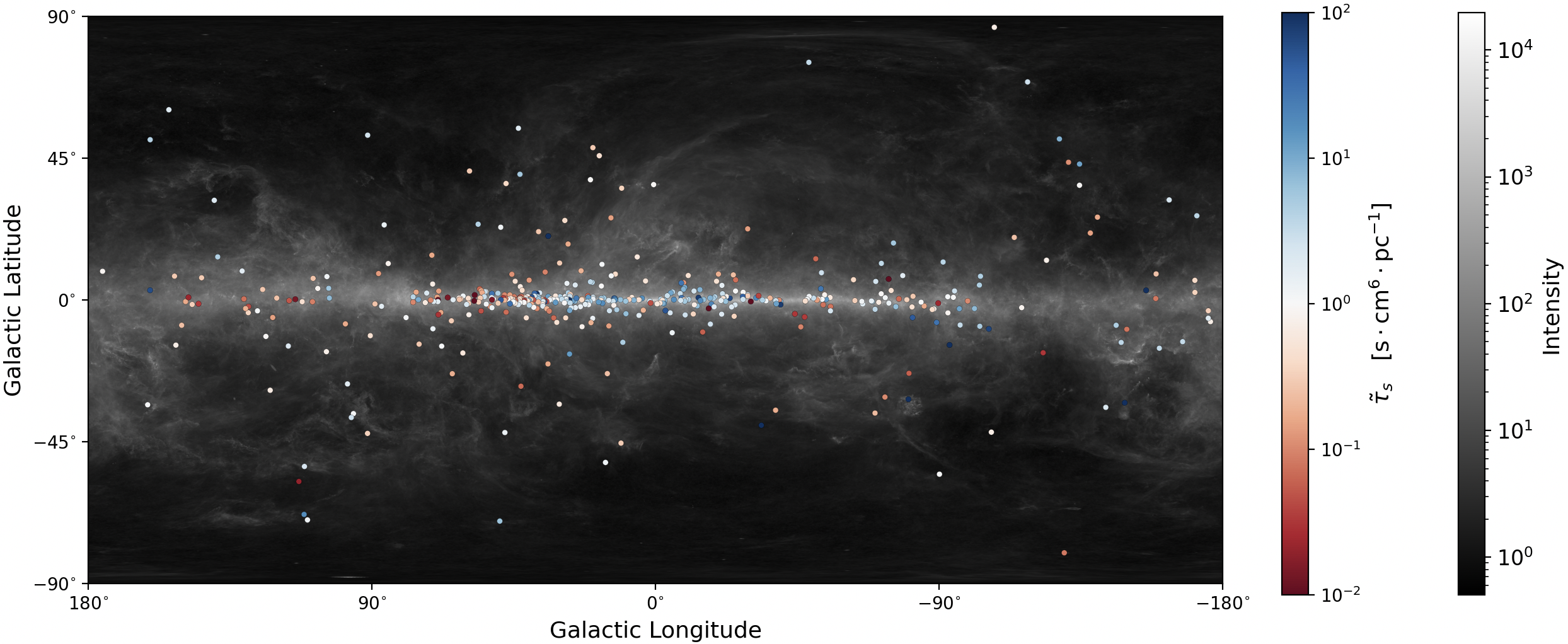}
    \includegraphics[width=\linewidth]{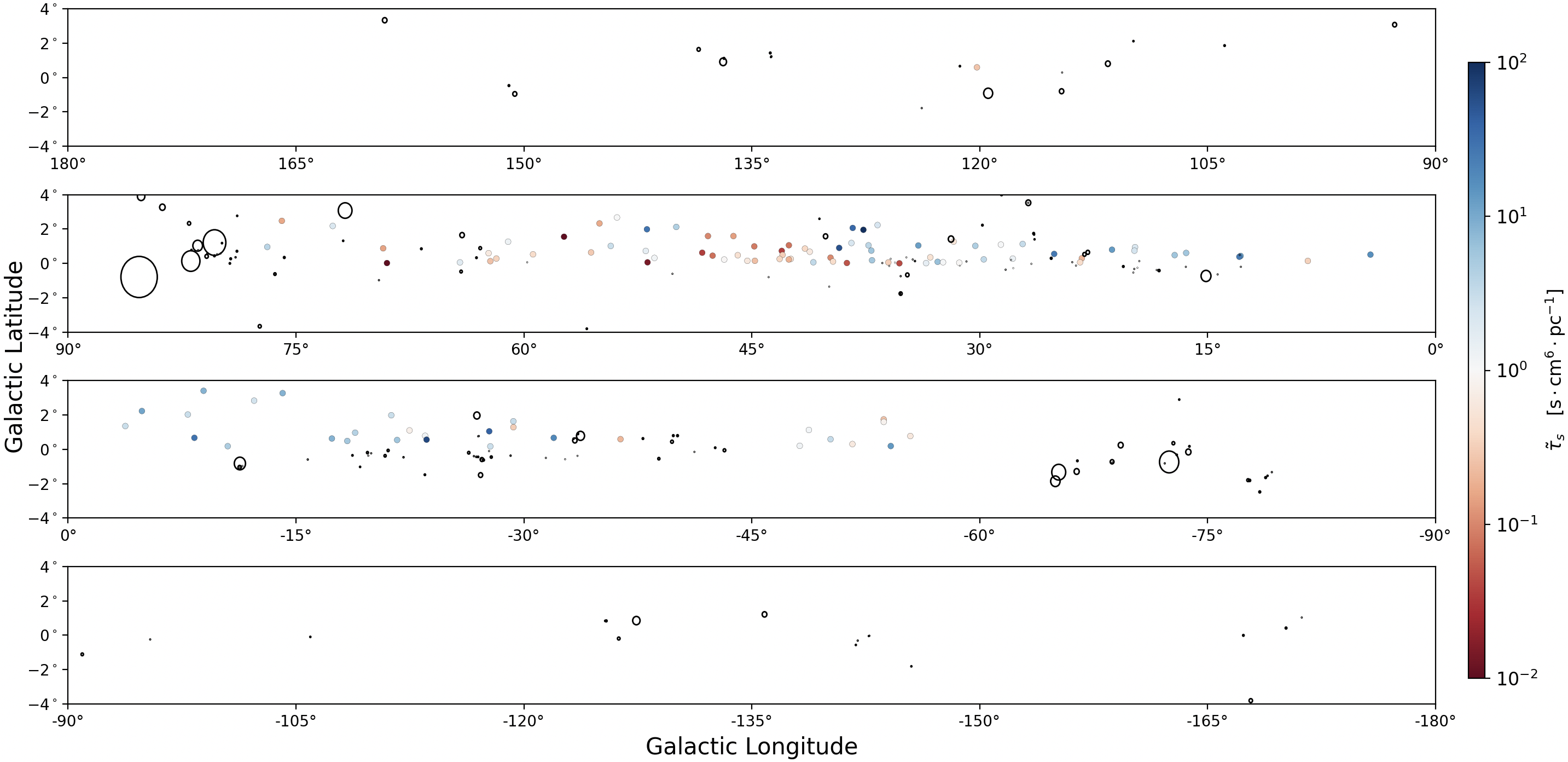}
    \caption{\textbf{Distribution of pulsar fluctuation scattering intensity ($\tilde{\tau}_{\rm s}$) in the Galaxy in comparison with other ISM data.} \textbf{Top}: comparison with the Planck 857\,GHz map. \textbf{Bottom}: comparison with the ${\rm H}\,_{\rm II}$ regions (black circles) identified with the WISE data.}
    \label{fig:all_data}
\end{figure*}


\bibliography{sample631}

\begin{thebibliography}{}
\expandafter\ifx\csname natexlab\endcsname\relax\def\natexlab#1{#1}\fi
\providecommand{\url}[1]{\href{#1}{#1}}
\providecommand{\dodoi}[1]{doi:~\href{http://doi.org/#1}{\nolinkurl{#1}}}
\providecommand{\doeprint}[1]{\href{http://ascl.net/#1}{\nolinkurl{http://ascl.net/#1}}}
\providecommand{\doarXiv}[1]{\href{https://arxiv.org/abs/#1}{\nolinkurl{https://arxiv.org/abs/#1}}}

\bibitem[{{Alexander} {et~al.}(1971){Alexander}, {Brandt}, {Maran}, \& {Stecher}}]{Gumtem2}
{Alexander}, J.~K., {Brandt}, J.~C., {Maran}, S.~P., \& {Stecher}, T.~P. 1971, \apj, 167, 487, \dodoi{10.1086/151045}

\bibitem[{{Anderson} {et~al.}(2014){Anderson}, {Bania}, {Balser}, {Cunningham}, {Wenger}, {Johnstone}, \& {Armentrout}}]{WISE}
{Anderson}, L.~D., {Bania}, T.~M., {Balser}, D.~S., {et~al.} 2014, \apjs, 212, 1, \dodoi{10.1088/0067-0049/212/1/1}

\bibitem[{{Armstrong}(1984)}]{scint2}
{Armstrong}, J.~W. 1984, \nat, 307, 527, \dodoi{10.1038/307527a0}

\bibitem[{{Backer}(1974)}]{Vela0}
{Backer}, D.~C. 1974, \apj, 190, 667, \dodoi{10.1086/152924}

\bibitem[{{Baker} {et~al.}(2023){Baker}, {Brisken}, {van Kerkwijk}, {van Lieshout}, \& {Pen}}]{Baker23}
{Baker}, D., {Brisken}, W., {van Kerkwijk}, M.~H., {van Lieshout}, R., \& {Pen}, U.-L. 2023, \mnras, 525, 211, \dodoi{10.1093/mnras/stad2318}

\bibitem[{{Beuermann}(1973)}]{Gumtem1}
{Beuermann}, K.~P. 1973, \apss, 20, 27, \dodoi{10.1007/BF00645581}

\bibitem[{{Bhat} {et~al.}(2004){Bhat}, {Cordes}, {Camilo}, {Nice}, \& {Lorimer}}]{17.2004ApJ...605..759B}
{Bhat}, N.~D.~R., {Cordes}, J.~M., {Camilo}, F., {Nice}, D.~J., \& {Lorimer}, D.~R. 2004, \apj, 605, 759, \dodoi{10.1086/382680}

\bibitem[{{Brisken} {et~al.}(2010){Brisken}, {Macquart}, {Gao}, {Rickett}, {Coles}, {Deller}, {Tingay}, \& {West}}]{Brisken10}
{Brisken}, W.~F., {Macquart}, J.~P., {Gao}, J.~J., {et~al.} 2010, \apj, 708, 232, \dodoi{10.1088/0004-637X/708/1/232}

\bibitem[{{Bruhweiler} {et~al.}(1979){Bruhweiler}, {Kondo}, \& {McCluskey}}]{Gumtem5}
{Bruhweiler}, F.~C., {Kondo}, J., \& {McCluskey}, G.~E., J. 1979, \apjl, 229, L39, \dodoi{10.1086/182926}

\bibitem[{{Cane}(1973)}]{Gumtem4}
{Cane}, H.~V. 1973, \pasa, 2, 197, \dodoi{10.1017/S1323358000013540}

\bibitem[{{Cordes} {et~al.}(2004){Cordes}, {Bhat}, {Hankins}, {McLaughlin}, \& {Kern}}]{cxx1}
{Cordes}, J.~M., {Bhat}, N.~D.~R., {Hankins}, T.~H., {McLaughlin}, M.~A., \& {Kern}, J. 2004, \apj, 612, 375, \dodoi{10.1086/422495}

\bibitem[{{Cordes} \& {Lazio}(2002)}]{Cordes02}
{Cordes}, J.~M., \& {Lazio}, T.~J.~W. 2002, arXiv e-prints, astro, \dodoi{10.48550/arXiv.astro-ph/0207156}

\bibitem[{{Cordes} \& {Lazio}(2003)}]{Cordes03}
---. 2003, arXiv e-prints, astro, \dodoi{10.48550/arXiv.astro-ph/0301598}

\bibitem[{{Cordes} {et~al.}(2006){Cordes}, {Rickett}, {Stinebring}, \& {Coles}}]{Cordes06}
{Cordes}, J.~M., {Rickett}, B.~J., {Stinebring}, D.~R., \& {Coles}, W.~A. 2006, \apj, 637, 346, \dodoi{10.1086/498332}

\bibitem[{{Davidson} \& {Terzian}(1969)}]{Davidson69}
{Davidson}, K., \& {Terzian}, Y. 1969, \nat, 221, 729, \dodoi{10.1038/221729b0}

\bibitem[{{Dennison} {et~al.}(1999){Dennison}, {Simonetti}, \& {Topasna}}]{VTSS}
{Dennison}, B., {Simonetti}, J.~H., \& {Topasna}, G.~A. 1999, in American Astronomical Society Meeting Abstracts, Vol. 195, American Astronomical Society Meeting Abstracts, 53.09

\bibitem[{{Deshpande} {et~al.}(1992){Deshpande}, {McCulloch}, {Radhakrishnan}, \& {Anantharamaiah}}]{HI}
{Deshpande}, A.~A., {McCulloch}, P.~M., {Radhakrishnan}, V., \& {Anantharamaiah}, K.~R. 1992, \mnras, 258, 19P, \dodoi{10.1093/mnras/258.1.19P}

\bibitem[{{Duncan} {et~al.}(1996){Duncan}, {Stewart}, {Haynes}, \& {Jones}}]{Vela}
{Duncan}, A.~R., {Stewart}, R.~T., {Haynes}, R.~F., \& {Jones}, K.~L. 1996, \mnras, 280, 252, \dodoi{10.1093/mnras/280.1.252}

\bibitem[{{Egger}(1998)}]{LB}
{Egger}, R. 1998, in IAU Colloq. 166: The Local Bubble and Beyond, ed. D.~{Breitschwerdt}, M.~J. {Freyberg}, \& J.~{Truemper}, Vol. 506, 287--296, \dodoi{10.1007/BFb0104735}

\bibitem[{{Ellis}(1972)}]{Gumtem3}
{Ellis}, G.~R.~A. 1972, \pasa, 2, 158, \dodoi{10.1017/S1323358000013412}

\bibitem[{{Finkbeiner}(2003)}]{Halpha}
{Finkbeiner}, D.~P. 2003, \apjs, 146, 407, \dodoi{10.1086/374411}

\bibitem[{{Gaustad} {et~al.}(2001){Gaustad}, {McCullough}, {Rosing}, \& {Van Buren}}]{SHASSA}
{Gaustad}, J.~E., {McCullough}, P.~R., {Rosing}, W., \& {Van Buren}, D. 2001, \pasp, 113, 1326, \dodoi{10.1086/323969}

\bibitem[{{Gum}(1952)}]{Gum52}
{Gum}, C.~S. 1952, The Observatory, 72, 151

\bibitem[{{Han} {et~al.}(2021){Han}, {Wang}, {Wang}, {Wang}, {Zhou}, {Sun}, {Yan}, {Su}, {Jing}, {Chen}, {Gao}, {Hou}, {Xu}, {Lee}, {Wang}, {Jiang}, {Xu}, {Yan}, {Gan}, {Guan}, {Huang}, {Jiang}, {Li}, {Men}, {Sun}, {Wang}, {Wang}, {Wang}, {Xie}, {Xu}, {Yao}, {You}, {Yu}, {Yuan}, {Yuen}, {Zhang}, \& {Zhu}}]{GPPS}
{Han}, J.~L., {Wang}, C., {Wang}, P.~F., {et~al.} 2021, Research in Astronomy and Astrophysics, 21, 107, \dodoi{10.1088/1674-4527/21/5/107}

\bibitem[{{He} \& {Shi}(2024)}]{He24}
{He}, Q., \& {Shi}, X. 2024, \mnras, 527, 5183, \dodoi{10.1093/mnras/stad3561}

\bibitem[{{HI4PI Collaboration} {et~al.}(2016){HI4PI Collaboration}, {Ben Bekhti}, {Fl{\"o}er}, {Keller}, {Kerp}, {Lenz}, {Winkel}, {Bailin}, {Calabretta}, {Dedes}, {Ford}, {Gibson}, {Haud}, {Janowiecki}, {Kalberla}, {Lockman}, {McClure-Griffiths}, {Murphy}, {Nakanishi}, {Pisano}, \& {Staveley-Smith}}]{HI4PI}
{HI4PI Collaboration}, {Ben Bekhti}, N., {Fl{\"o}er}, L., {et~al.} 2016, \aap, 594, A116, \dodoi{10.1051/0004-6361/201629178}

\bibitem[{{Hill} {et~al.}(2005){Hill}, {Stinebring}, {Asplund}, {Berwick}, {Everett}, \& {Hinkel}}]{Hill05}
{Hill}, A.~S., {Stinebring}, D.~R., {Asplund}, C.~T., {et~al.} 2005, \apjl, 619, L171, \dodoi{10.1086/428347}

\bibitem[{{Hill} {et~al.}(2003){Hill}, {Stinebring}, {Barnor}, {Berwick}, \& {Webber}}]{Hill03}
{Hill}, A.~S., {Stinebring}, D.~R., {Barnor}, H.~A., {Berwick}, D.~E., \& {Webber}, A.~B. 2003, \apj, 599, 457, \dodoi{10.1086/379191}

\bibitem[{{Hou} {et~al.}(2022){Hou}, {Han}, {Hong}, {Gao}, \& {Wang}}]{Hou2}
{Hou}, L., {Han}, J., {Hong}, T., {Gao}, X., \& {Wang}, C. 2022, Science China Physics, Mechanics, and Astronomy, 65, 129703, \dodoi{10.1007/s11433-022-2039-8}

\bibitem[{{Johnston} {et~al.}(1998){Johnston}, {Nicastro}, \& {Koribalski}}]{Vela2}
{Johnston}, S., {Nicastro}, L., \& {Koribalski}, B. 1998, \mnras, 297, 108, \dodoi{10.1046/j.1365-8711.1998.01461.x}

\bibitem[{{Kalberla} {et~al.}(2010){Kalberla}, {McClure-Griffiths}, {Pisano}, {Calabretta}, {Ford}, {Lockman}, {Staveley-Smith}, {Kerp}, {Winkel}, {Murphy}, \& {Newton-McGee}}]{HI10}
{Kalberla}, P.~M.~W., {McClure-Griffiths}, N.~M., {Pisano}, D.~J., {et~al.} 2010, \aap, 521, A17, \dodoi{10.1051/0004-6361/200913979}

\bibitem[{{Kerp} {et~al.}(2016){Kerp}, {Kalberla}, {Ben Bekhti}, {Fl{\"o}er}, {Lenz}, \& {Winkel}}]{HI16}
{Kerp}, J., {Kalberla}, P.~M.~W., {Ben Bekhti}, N., {et~al.} 2016, \aap, 589, A120, \dodoi{10.1051/0004-6361/201526395}

\bibitem[{{Lallement} {et~al.}(2019){Lallement}, {Babusiaux}, {Vergely}, {Katz}, {Arenou}, {Valette}, {Hottier}, \& {Capitanio}}]{dustmap}
{Lallement}, R., {Babusiaux}, C., {Vergely}, J.~L., {et~al.} 2019, \aap, 625, A135, \dodoi{10.1051/0004-6361/201834695}

\bibitem[{{Lee} \& {Jokipii}(1975)}]{scint1}
{Lee}, L.~C., \& {Jokipii}, J.~R. 1975, \apj, 201, 532, \dodoi{10.1086/153916}

\bibitem[{{Lerche}(1970)}]{dispersion1}
{Lerche}, I. 1970, \apss, 6, 287, \dodoi{10.1007/BF00651229}

\bibitem[{{Lyne} {et~al.}(1985){Lyne}, {Manchester}, \& {Taylor}}]{Lyne85}
{Lyne}, A.~G., {Manchester}, R.~N., \& {Taylor}, J.~H. 1985, \mnras, 213, 613, \dodoi{10.1093/mnras/213.3.613}

\bibitem[{{Manchester} {et~al.}(2005){Manchester}, {Hobbs}, {Teoh}, \& {Hobbs}}]{ATNF}
{Manchester}, R.~N., {Hobbs}, G.~B., {Teoh}, A., \& {Hobbs}, M. 2005, \aj, 129, 1993, \dodoi{10.1086/428488}

\bibitem[{{Manchester} \& {Taylor}(1981)}]{Manchester81}
{Manchester}, R.~N., \& {Taylor}, J.~H. 1981, \aj, 86, 1953, \dodoi{10.1086/113078}

\bibitem[{{McKee} \& {Williams}(1997)}]{HII02}
{McKee}, C.~F., \& {Williams}, J.~P. 1997, \apj, 476, 144, \dodoi{10.1086/303587}

\bibitem[{{McKee} {et~al.}(2022){McKee}, {Zhu}, {Stinebring}, \& {Cordes}}]{B1133}
{McKee}, J.~W., {Zhu}, H., {Stinebring}, D.~R., \& {Cordes}, J.~M. 2022, \apj, 927, 99, \dodoi{10.3847/1538-4357/ac460b}

\bibitem[{{Mitra} \& {Ramachandran}(2001)}]{Vela4}
{Mitra}, D., \& {Ramachandran}, R. 2001, \aap, 370, 586, \dodoi{10.1051/0004-6361:20010274}

\bibitem[{{Mohan} {et~al.}(2002){Mohan}, {Anantharamaiah}, \& {Goss}}]{RRL1}
{Mohan}, N.~R., {Anantharamaiah}, K.~R., \& {Goss}, W.~M. 2002, \apj, 574, 701, \dodoi{10.1086/341004}

\bibitem[{{Morini}(1983)}]{Vela1}
{Morini}, M. 1983, \mnras, 202, 495, \dodoi{10.1093/mnras/202.2.495}

\bibitem[{{Ocker} {et~al.}(2023){Ocker}, {Cordes}, \& {Chatterjee}}]{Ocker2}
{Ocker}, S., {Cordes}, J., \& {Chatterjee}, S. 2023, in AGU Fall Meeting Abstracts, Vol. 2023, SH43D--319

\bibitem[{{Ocker} {et~al.}(2024{\natexlab{a}}){Ocker}, {Anderson}, {Lazio}, {Cordes}, \& {Ravi}}]{Ocker24}
{Ocker}, S.~K., {Anderson}, L.~D., {Lazio}, T. J.~W., {Cordes}, J.~M., \& {Ravi}, V. 2024{\natexlab{a}}, \apj, 974, 10, \dodoi{10.3847/1538-4357/ad6a51}

\bibitem[{{Ocker} {et~al.}(2024{\natexlab{b}}){Ocker}, {Cordes}, {Chatterjee}, {Stinebring}, {Dolch}, {Giannakopoulos}, {Pelgrims}, {McKee}, \& {Reardon}}]{Ocker}
{Ocker}, S.~K., {Cordes}, J.~M., {Chatterjee}, S., {et~al.} 2024{\natexlab{b}}, \mnras, 527, 7568, \dodoi{10.1093/mnras/stad3683}

\bibitem[{{Oswald} {et~al.}(2021){Oswald}, {Karastergiou}, {Posselt}, {Johnston}, {Bailes}, {Buchner}, {Geyer}, {Keith}, {Kramer}, {Parthasarathy}, {Reardon}, {Serylak}, {Shannon}, {Spiewak}, {van Straten}, \& {Venkatraman Krishnan}}]{58.2021MNRAS.504.1115O}
{Oswald}, L.~S., {Karastergiou}, A., {Posselt}, B., {et~al.} 2021, \mnras, 504, 1115, \dodoi{10.1093/mnras/stab980}

\bibitem[{{Pen} \& {Levin}(2014)}]{Pen14}
{Pen}, U.-L., \& {Levin}, Y. 2014, \mnras, 442, 3338, \dodoi{10.1093/mnras/stu1020}

\bibitem[{{Planck Collaboration} {et~al.}(2020{\natexlab{a}}){Planck Collaboration}, {Aghanim}, {Akrami}, {Arroja}, {Ashdown}, {Aumont}, {Baccigalupi}, {Ballardini}, {Banday}, {Barreiro}, {Bartolo}, {Basak}, {Battye}, {Benabed}, {Bernard}, {Bersanelli}, {Bielewicz}, {Bock}, {Bond}, {Borrill}, {Bouchet}, {Boulanger}, {Bucher}, {Burigana}, {Butler}, {Calabrese}, {Cardoso}, {Carron}, {Casaponsa}, {Challinor}, {Chiang}, {Colombo}, {Combet}, {Contreras}, {Crill}, {Cuttaia}, {de Bernardis}, {de Zotti}, {Delabrouille}, {Delouis}, {D{\'e}sert}, {Di Valentino}, {Dickinson}, {Diego}, {Donzelli}, {Dor{\'e}}, {Douspis}, {Ducout}, {Dupac}, {Efstathiou}, {Elsner}, {En{\ss}lin}, {Eriksen}, {Falgarone}, {Fantaye}, {Fergusson}, {Fernandez-Cobos}, {Finelli}, {Forastieri}, {Frailis}, {Franceschi}, {Frolov}, {Galeotta}, {Galli}, {Ganga}, {G{\'e}nova-Santos}, {Gerbino}, {Ghosh}, {Gonz{\'a}lez-Nuevo}, {G{\'o}rski}, {Gratton}, {Gruppuso}, {Gudmundsson}, {Hamann}, {Handley}, {Hansen}, {Helou}, {Herranz}, {Hildebrandt}, {Hivon}, {Huang}, {Jaffe}, {Jones}, {Karakci}, {Keih{\"a}nen}, {Keskitalo}, {Kiiveri}, {Kim}, {Kisner}, {Knox}, {Krachmalnicoff}, {Kunz}, {Kurki-Suonio}, {Lagache}, {Lamarre}, {Langer}, {Lasenby}, {Lattanzi}, {Lawrence}, {Le Jeune}, {Leahy}, {Lesgourgues}, {Levrier}, {Lewis}, {Liguori}, {Lilje}, {Lilley}, {Lindholm}, {L{\'o}pez-Caniego}, {Lubin}, {Ma}, {Mac{\'\i}as-P{\'e}rez}, {Maggio}, {Maino}, {Mandolesi}, {Mangilli}, {Marcos-Caballero}, {Maris}, {Martin}, {Martinelli}, {Mart{\'\i}nez-Gonz{\'a}lez}, {Matarrese}, {Mauri}, {McEwen}, {Meerburg}, {Meinhold}, {Melchiorri}, {Mennella}, {Migliaccio}, {Millea}, {Mitra}, {Miville-Desch{\^e}nes}, {Molinari}, {Moneti}, {Montier}, {Morgante}, {Moss}, {Mottet}, {M{\"u}nchmeyer}, {Natoli}, {N{\o}rgaard-Nielsen}, {Oxborrow}, {Pagano}, {Paoletti}, {Partridge}, {Patanchon}, {Pearson}, {Peel}, {Peiris}, {Perrotta}, {Pettorino}, {Piacentini}, {Polastri}, {Polenta}, {Puget}, {Rachen}, {Reinecke}, {Remazeilles}, {Renault}, {Renzi}, {Rocha}, {Rosset}, {Roudier}, {Rubi{\~n}o-Mart{\'\i}n},
  {Ruiz-Granados}, {Salvati}, {Sandri}, {Savelainen}, {Scott}, {Shellard}, {Shiraishi}, {Sirignano}, {Sirri}, {Spencer}, {Sunyaev}, {Suur-Uski}, {Tauber}, {Tavagnacco}, {Tenti}, {Terenzi}, {Toffolatti}, {Tomasi}, {Trombetti}, {Valiviita}, {Van Tent}, {Vibert}, {Vielva}, {Villa}, {Vittorio}, {Wandelt}, {Wehus}, {White}, {White}, {Zacchei}, \& {Zonca}}]{Plank19}
{Planck Collaboration}, {Aghanim}, N., {Akrami}, Y., {et~al.} 2020{\natexlab{a}}, \aap, 641, A1, \dodoi{10.1051/0004-6361/201833880}

\bibitem[{{Planck Collaboration} {et~al.}(2020{\natexlab{b}}){Planck Collaboration}, {Akrami}, {Arg{\"u}eso}, {Ashdown}, {Aumont}, {Baccigalupi}, {Ballardini}, {Banday}, {Barreiro}, {Bartolo}, {Basak}, {Benabed}, {Bernard}, {Bersanelli}, {Bielewicz}, {Bonavera}, {Bond}, {Borrill}, {Bouchet}, {Boulanger}, {Bucher}, {Burigana}, {Butler}, {Calabrese}, {Cardoso}, {Colombo}, {Crill}, {Cuttaia}, {de Bernardis}, {de Rosa}, {de Zotti}, {Delabrouille}, {Di Valentino}, {Dickinson}, {Diego}, {Donzelli}, {Ducout}, {Dupac}, {Efstathiou}, {Elsner}, {En{\ss}lin}, {Eriksen}, {Fantaye}, {Finelli}, {Frailis}, {Franceschi}, {Frolov}, {Galeotta}, {Galli}, {Ganga}, {G{\'e}nova-Santos}, {Gerbino}, {Ghosh}, {Gonz{\'a}lez-Nuevo}, {G{\'o}rski}, {Gratton}, {Gruppuso}, {Gudmundsson}, {Handley}, {Hansen}, {Herranz}, {Hivon}, {Huang}, {Jaffe}, {Jones}, {Karakci}, {Keih{\"a}nen}, {Keskitalo}, {Kiiveri}, {Kim}, {Kisner}, {Krachmalnicoff}, {Kunz}, {Kurki-Suonio}, {Lamarre}, {Lasenby}, {Lattanzi}, {Lawrence}, {Leahy}, {Levrier}, {Liguori}, {Lilje}, {Lindholm}, {L{\'o}pez-Caniego}, {Ma}, {Mac{\'\i}as-P{\'e}rez}, {Maggio}, {Maino}, {Mandolesi}, {Mangilli}, {Maris}, {Martin}, {Mart{\'\i}nez-Gonz{\'a}lez}, {Matarrese}, {Mauri}, {McEwen}, {Meinhold}, {Melchiorri}, {Mennella}, {Migliaccio}, {Molinari}, {Montier}, {Morgante}, {Moss}, {Natoli}, {Pagano}, {Paoletti}, {Partridge}, {Patanchon}, {Patrizii}, {Peel}, {Perrotta}, {Pettorino}, {Piacentini}, {Polenta}, {Puget}, {Rachen}, {Racine}, {Reinecke}, {Remazeilles}, {Renzi}, {Rocha}, {Roudier}, {Rubi{\~n}o-Mart{\'\i}n}, {Salvati}, {Sandri}, {Savelainen}, {Scott}, {Seljebotn}, {Sirignano}, {Sirri}, {Spencer}, {Suur-Uski}, {Tauber}, {Tavagnacco}, {Tenti}, {Terenzi}, {Toffolatti}, {Tomasi}, {Trombetti}, {Valiviita}, {Vansyngel}, {Van Tent}, {Vielva}, {Villa}, {Vittorio}, {Wandelt}, {Watson}, {Wehus}, {Zacchei}, \& {Zonca}}]{Plank192}
{Planck Collaboration}, {Akrami}, Y., {Arg{\"u}eso}, F., {et~al.} 2020{\natexlab{b}}, \aap, 641, A2, \dodoi{10.1051/0004-6361/201833293}

\bibitem[{{Prentice} \& {Ter Haar}(1969)}]{HII69}
{Prentice}, A.~J.~R., \& {Ter Haar}, D. 1969, \nat, 222, 964, \dodoi{10.1038/222964a0}

\bibitem[{{Quireza} {et~al.}(2006){Quireza}, {Rood}, {Balser}, \& {Bania}}]{RRL2}
{Quireza}, C., {Rood}, R.~T., {Balser}, D.~S., \& {Bania}, T.~M. 2006, \apjs, 165, 338, \dodoi{10.1086/503901}

\bibitem[{{Reid} {et~al.}(2016){Reid}, {Dame}, {Menten}, \& {Brunthaler}}]{v_d16}
{Reid}, M.~J., {Dame}, T.~M., {Menten}, K.~M., \& {Brunthaler}, A. 2016, \apj, 823, 77, \dodoi{10.3847/0004-637X/823/2/77}

\bibitem[{{Reid} {et~al.}(2014){Reid}, {Menten}, {Brunthaler}, {Zheng}, {Dame}, {Xu}, {Wu}, {Zhang}, {Sanna}, {Sato}, {Hachisuka}, {Choi}, {Immer}, {Moscadelli}, {Rygl}, \& {Bartkiewicz}}]{v_d14}
{Reid}, M.~J., {Menten}, K.~M., {Brunthaler}, A., {et~al.} 2014, \apj, 783, 130, \dodoi{10.1088/0004-637X/783/2/130}

\bibitem[{{Reid} {et~al.}(2019){Reid}, {Menten}, {Brunthaler}, {Zheng}, {Dame}, {Xu}, {Li}, {Sakai}, {Wu}, {Immer}, {Zhang}, {Sanna}, {Moscadelli}, {Rygl}, {Bartkiewicz}, {Hu}, {Quiroga-Nu{\~n}ez}, \& {van Langevelde}}]{v_d19}
---. 2019, \apj, 885, 131, \dodoi{10.3847/1538-4357/ab4a11}

\bibitem[{{Reynolds} {et~al.}(2002){Reynolds}, {Haffner}, \& {Madsen}}]{WHAM}
{Reynolds}, R.~J., {Haffner}, L.~M., \& {Madsen}, G.~J. 2002, in Astronomical Society of the Pacific Conference Series, Vol. 282, Galaxies: the Third Dimension, ed. M.~{Rosada}, L.~{Binette}, \& L.~{Arias}, 31, \dodoi{10.48550/arXiv.astro-ph/0201392}

\bibitem[{{Rickett}(1970)}]{Rickett70}
{Rickett}, B.~J. 1970, \mnras, 150, 67, \dodoi{10.1093/mnras/150.1.67}

\bibitem[{{Rickett}(1975)}]{dispersion2}
---. 1975, \apj, 197, 185, \dodoi{10.1086/153501}

\bibitem[{{Scheuer}(1968)}]{Scheuer68}
{Scheuer}, P.~A.~G. 1968, \nat, 218, 920, \dodoi{10.1038/218920a0}

\bibitem[{{Shi}(2021)}]{SX}
{Shi}, X. 2021, \mnras, 508, 125, \dodoi{10.1093/mnras/stab2522}

\bibitem[{{Spangler}(1991)}]{HII01}
{Spangler}, S.~R. 1991, \apj, 376, 540, \dodoi{10.1086/170303}

\bibitem[{{Stinebring} {et~al.}(2001){Stinebring}, {McLaughlin}, {Cordes}, {Becker}, {Goodman}, {Kramer}, {Sheckard}, \& {Smith}}]{Stinebring01}
{Stinebring}, D.~R., {McLaughlin}, M.~A., {Cordes}, J.~M., {et~al.} 2001, \apjl, 549, L97, \dodoi{10.1086/319133}

\bibitem[{{Stinebring} {et~al.}(2000){Stinebring}, {Smirnova}, {Hankins}, {Hovis}, {Kaspi}, {Kempner}, {Myers}, \& {Nice}}]{Vela3}
{Stinebring}, D.~R., {Smirnova}, T.~V., {Hankins}, T.~H., {et~al.} 2000, \apj, 539, 300, \dodoi{10.1086/309201}

\bibitem[{{Stock} \& {van Kerkwijk}(2024)}]{Stock24}
{Stock}, A.~M., \& {van Kerkwijk}, M.~H. 2024, arXiv e-prints, arXiv:2407.16876, \dodoi{10.48550/arXiv.2407.16876}

\bibitem[{{Sutton}(1971)}]{Sutton71}
{Sutton}, J.~M. 1971, \mnras, 155, 51, \dodoi{10.1093/mnras/155.1.51}

\bibitem[{{Taylor} \& {Cordes}(1993)}]{Taylor93}
{Taylor}, J.~H., \& {Cordes}, J.~M. 1993, \apj, 411, 674, \dodoi{10.1086/172870}

\bibitem[{{Walker} {et~al.}(2004){Walker}, {Melrose}, {Stinebring}, \& {Zhang}}]{Walker04}
{Walker}, M.~A., {Melrose}, D.~B., {Stinebring}, D.~R., \& {Zhang}, C.~M. 2004, \mnras, 354, 43, \dodoi{10.1111/j.1365-2966.2004.08159.x}

\bibitem[{{Walker} {et~al.}(2017){Walker}, {Tuntsov}, {Bignall}, {Reynolds}, {Bannister}, {Johnston}, {Stevens}, \& {Ravi}}]{walker17}
{Walker}, M.~A., {Tuntsov}, A.~V., {Bignall}, H., {et~al.} 2017, \apj, 843, 15, \dodoi{10.3847/1538-4357/aa705c}

\bibitem[{{Xu} {et~al.}(2023){Xu}, {Shi}, {Lee}, {Hao}, {Li}, {Wang}, {Yuan}, {Xu}, {Wu}, {Jiang}, {Huang}, {Wang}, {Shen}, \& {Cao}}]{Xu_Vela}
{Xu}, Y., {Shi}, X., {Lee}, K., {et~al.} 2023, \mnras, 526, 1246, \dodoi{10.1093/mnras/stad2837}

\bibitem[{{Yao} {et~al.}(2021){Yao}, {Zhu}, {Manchester}, {Coles}, {Li}, {Wang}, {Kramer}, {Stinebring}, {Feng}, {Yan}, {Miao}, {Yuan}, {Wang}, \& {Lu}}]{cxx2}
{Yao}, J., {Zhu}, W., {Manchester}, R.~N., {et~al.} 2021, Nature Astronomy, 5, 788, \dodoi{10.1038/s41550-021-01360-w}

\bibitem[{{Yao} {et~al.}(2017){Yao}, {Manchester}, \& {Wang}}]{YMW16}
{Yao}, J.~M., {Manchester}, R.~N., \& {Wang}, N. 2017, \apj, 835, 29, \dodoi{10.3847/1538-4357/835/1/29}

\end{thebibliography}
\bibliographystyle{aasjournal}



\end{document}